\documentclass[showpacs,floatfix,preprint,nofootinbib,longbibliography]{revtex4-1}
\usepackage{graphicx,psfrag,amsmath,amssymb,amsfonts,latexsym,color,graphpap,dcolumn}
\usepackage{bbm}
\usepackage{bm}
\usepackage[caption=false]{subfig}
\usepackage{caption}

\usepackage{hyperref}
\hypersetup{colorlinks=true,linktoc=page,linkcolor=blue,citecolor=red,urlcolor=cyan}

\newcommand{\colm}[1]{\textcolor{magenta}{#1}}
\newcommand{\colred}[1]{\textcolor{red}{(\sc #1)}}

\usepackage[english]{babel}

\newcommand{\hu}{\hspace{1cm}}
\newcommand{\alls}[1][x]{\mathbf{#1}}
\newcommand{\paras}[1][x]{\bm{#1}^\parallel}
\newcommand{\para}[1][x]{{{#1}^{\parallel}}}
\newcommand{\perpe}[1][x]{{#1}^3}
\newcommand{\perpez}[1][x]{z}
\newcommand{\dime}{D}
\newcommand{\dk}[1][k]{\frac{{\rm d} #1}{(2\pi)}}
\newcommand{\dx}[1][x]{{{\rm d} #1}}

\newcommand{\mathi}{{\text{i}}}

\newcommand{\basicF}{F}

\newcommand{\ren}[1]{\bigl\langle #1 \bigr\rangle_{\rm ren}}
\newcommand{\vev}[1]{\bigl\langle #1 \bigr\rangle}

\newcommand{{\se}}{{SE}}
\usepackage{xparse}

\NewDocumentCommand{\dkpara}{O{k} O{\dime}}{\frac{{\rm d} {#1}^{\parallel}}{(2\pi)^{#2}}}
\NewDocumentCommand{\dkd}{O{k} O{\dime}}{ \frac{{\rm d}^{#2} {#1}}{(2\pi)^{#2}}}
\NewDocumentCommand{\dxd}{O{x} O{\dime}}{ {\rm d}^{#2} {#1} }
\NewDocumentCommand{\ad}{O{} O{}}{ \bigl \langle {#1}\bigr\rangle_{{\rm ad}{#2}} }

\newcommand{\measure}[1]{\dxd[{#1}_1][4] \cdots \dxd[{#1}_n][4]}

\renewcommand{\vec}[1]{\alls[#1]}

\begin{document}
\title{Quantum vacuum fluctuations and the principle of virtual work in inhomogeneous backgrounds}
\author{S.~A.~Franchino-Vi\~nas$^{1,2}$, M.~N.~Manti\~nan$^{3,4}$  and F.~D.~Mazzitelli $^{3,4}$}
\affiliation{$^1$Departamento de F\'isica, Facultad de Ciencias Exactas\\
Universidad Nacional de La Plata, C.C.\ 67 (1900), La Plata, Argentina\\
 $^2$ Institut f\"ur Theoretische Physik, Universit\"at Heidelberg\\ D-69120 Heidelberg, Germany.\\
$^3$Centro At\'omico Bariloche,  
Comisi\'on Nacional de Energ\'\i a At\'omica, R8402AGP Bariloche, Argentina\\
$^4$ Instituto Balseiro, Universidad Nacional de Cuyo, R8402AGP Bariloche, Argentina. }

\begin{abstract} 
We discuss several aspects of the stress-energy tensor for a quantum scalar field in an inhomogeneous background, the latter being modeled by a variable mass. 
Using a perturbative approach, dimensional regularization and adiabatic subtraction, 
we present all-order formal expressions for the stress-energy tensor.
Importantly, we provide an explicit proof of the principle of virtual work for Casimir forces, taking advantage of the conservation law for the renormalized stress-energy tensor. 
We discuss also discontinuity-induced divergences. 
For the particular case of planar inhomogeneities, we corroborate the perturbative results 
 with a WKB-inspired expansion.
\end{abstract}
\date{\today}
\maketitle

\tableofcontents


\section{Introduction}\label{sec:intro}

Energy densities, stresses and forces are produced by vacuum fluctuations of the electromagnetic 
field when a body is immersed in a medium. In this context,
the immersion of bodies in homogeneous media was already considered in the seminal works \cite{Lifshitz:1956zz, Dzyaloshinskii_1961},  
where the so-called Lifshitz formula was derived. 
This formula is able to describe the force 
between two flat and parallel interphases that separate three different homogeneous media. 

More recently, there have been efforts to define the stress-energy (SE) tensor for
a quantum field in a generalized Lifshitz configuration, i.e. in a situation in which the media are characterized by spacetime-dependent electromagnetic properties.
Taking into account that
the renormalization originally proposed by Lifshitz et al. 
does not work in such a case, the problem 
has been considered by several authors  \cite{Bao:2015, Murray:2015tim, Griniasty:2017ofc, Fulling:2018qcn, Parashar:2018pds, Li:2019ohr, Shayit:2021kgn, Efrat:2021avl}.
Despite the different methods, models and  particular subtractions (at the level of either the Green functions or the {\se} tensor),
several questions are still open.

To discuss some of them, we will consider a toy model that consists of a quantum scalar field interacting with a classical field, in such a way that the quantum field acquires a variable mass. The evaluation of the vacuum expectation values (VEVs) will be performed using a perturbative approach in the variable mass. 
For the renormalization we will follow a standard approach, based on dimensional regularization and adiabatic subtraction. 

Observe first that 
similar theories have been analyzed in several contexts.
In particular much attention has been devoted to quantum fields in curved spacetimes \cite{Birrel:1982,Parker:2009}, 
for which there is a well-established procedure to obtain the renormalized 
{\se} tensor: infinities are absorbed into the bare constants of the theory. 
It is fairly obvious in this context that the renormalization
of the {\se} tensor's VEV
cannot be performed as suggested in \cite{Lifshitz:1956zz, Dzyaloshinskii_1961},  i.e.  by subtracting 
local quantities that depend only  on the value of the background fields at a given point: it must also involve 
derivatives of the background field. After absorbing the  divergences into the bare constants of the theory, 
the renormalized {\se} tensor will be expected to be defined up to local terms, which are determined by the finite part of the counterterms;
being local, they will not be relevant in the discussion of Casimir interactions between different bodies.

Notice that some general aspects of the renormalization procedure that we employ have been described in detail in Ref. \cite{Mazzitelli:2011st}. 
However, the calculations were performed to lowest order in the variable mass. To this order, it is possible to describe only the energy density and stresses, but not Casimir forces. Here we extend those results to arbitrary order. 
The case of a scalar field with variable mass depending on a single coordinate has also been considered in Ref. \cite{Fulling:2018qcn}, where the authors implement a Pauli--Villars renormalization along with a WKB subtraction. 
A similar approach, albeit with the scope of analyzing the limit of Dirichlet boundary conditions for thin surfaces, was followed in \cite{Graham:2002fw,Graham:2003ib,Franchino-Vinas:2020okl}.

One important aspect that we will discuss is that, if the background field models the presence of several bodies, the Casimir force can be then computed as customarily by taking the derivative of the system's vacuum energy  
with respect to the position of one of the bodies or, alternatively,   integrating the component of the {\se} tensor which is normal to the surface of that body. We will explicitly prove   that both approaches are equivalent, a result known as the principle of virtual work (PVW). 
In this respect, Ref.~\cite{Li:2019ohr} contains a discussion of the PVW for a planar configuration. 
Here we go beyond planar geometries; 
moreover, we provide an explicit connection with the PVW and the semiclassical conservation 
law of the {\se} tensor.

Previous works reported a ``pressure anomaly'', which may jeopardize the validity of the PVW \cite{Estrada:2012yn}.
It was later recognized that this anomaly is produced by a particular point-splitting regularization \cite{Murray:2015tim}.
Instead, our prescription using dimensional regularization along with adiabatic subtraction  guarantees the fulfillment of the conservation law 
and avoids the presence of anomalies.
In a recent work \cite{Efrat:2021avl} it has been pointed out
that quantum effects could induce a violation of the classical
relation between
the divergence of the electromagnetic stress and the 
gradients of the permeability and permittivity 
of the inhomogeneous media, inducing a ``van der Waals anomaly''. We have not found the analog of this anomaly in our
model. 

The last question that we will tackle is the fact that discontinuous backgrounds 
generate surface divergences in the VEV of the renormalized {\se} tensor. 
For the case of 
a perfectly conducting interphase, the presence of these divergences has been pointed out a long time ago \cite{Deutsch:1978sc}.
We will characterize this kind of divergences in a planar inhomogeneous model, and discuss 
its irrelevance in the calculation of Casimir forces. 
This  will be confirmed by nonperturbative calculations, based on a WKB-type approximation discussed in Refs. \cite{Bao:2015, Parashar:2018pds}  
for the case of
the electromagnetic field.



\medskip
The paper is organized as follows. 
In Section \ref{sec:model} we introduce our  model, which consists of a quantum scalar field $\phi$ in the presence of a background field $\sigma$ that provides an inhomogeneous mass term for the quantum field. 
In Section \ref{sec:ren} we discuss the renormalization of the VEVs $\vev{\phi^2}$ and
$\vev{T_{\mu\nu}^{(\phi)}}$, which is performed using standard techniques of quantum fields under the influence of external conditions. 
We also discuss the validity of the conservation law  of the renormalized  {\se} tensor at the semiclassical level. 
Section \ref{sec:perturbative} describes a perturbative approach for computing the abovementioned mean values, with particular emphasis in time-independent situations (i.e. when the background field is static). 
In Section \ref{sec:PVW} we prove the validity 
of the PVW.  The conservation law for the {\se} tensor turns out to be crucial in this context. 
Several examples are discussed then in Section \ref{sec:perturbative_examples}, including the surface divergences that appear in the renormalized mean values at the points where the background field is discontinuous. 
Afterwards, in Section \ref{sec:adiabatic} we reanalyze those surface divergences in the case of planar inhomogeneities within an adiabatic approach. 
Sec. \ref{sec:conc} contains the main conclusions of our work. 
Finally, the Apps. \ref{app:EM_order1}, \ref{app:divergent_EM_order2}, \ref{app:adiabatic_coefficients} and \ref{app:planar_boundaries} describe some further details of the calculations.

\bigskip
We use natural units $\hbar=c=1$ and metric signature $(+--\cdots)$ in a spacetime of dimension $\dime$. We define $g=-\det g_{\mu\nu}$
and spatial ($\dime-1$)-vectors are written in bold ($\alls[x]$).


\section{The model}\label{sec:model}

We will consider a quantum  field $\phi$ interacting with a background classical field $\sigma$ in the same fashion as in \cite{Mazzitelli:2011st}. 
The field $\sigma$ provides a variable mass for $\phi$,
so that the action for both fields on a curved background is given by \cite{Birrel:1982,Parker:2009}
\begin{equation}
\begin{split}
S =  \frac{1}{2}\int \dxd[x][\dime] \sqrt{g}\bigg[ & \phi_{,\mu} \phi^{,\mu}-\bigg(m_1^2+\xi_1 R +\frac{\lambda_1}{2}\sigma^2\bigg) \phi^2 
+ \sigma_{,\mu}\sigma^{,\mu}- (m_2^2+\xi_2 R )\sigma^2 - \frac{\lambda_2}{12} \sigma^4 \bigg]\, .
\end{split}
\label{eq: accion de materia}
\end{equation}
This theory can be considered as a toy model for the electromagnetic field in the presence of an inhomogeneous  medium. 
Of course, in order to mimic the electric permittivity or the magnetic permeability one could consider 
alternative models in which the coupling to the background field 
is through terms that involve spatial or time derivatives of $\phi$. 
However,  the action in Eq. \eqref{eq: accion de materia}  will be enough for our purposes. 

Even if we are interested in a four-dimensional spacetime, 
in Eq. \eqref{eq: accion de materia} we have introduced a dimensional regularization.
Moreover, the inclusion of a self-interacting term  for the background field, 
i.e. the one proportional to $\lambda_2$,  will be crucial for a succesful renormalization, 
as will also be the inclusion of a coupling to the curvature in curved spaces (terms proportional to $\xi_{1,2}$).
This will be discussed in detail in Sec. \ref{sec:ren}.

Peforming the variation of \eqref{eq: accion de materia} with respect to both fields, one can obtain the classical field equations, which read
\begin{align}
\left(\Box + m_1^2 + \xi_1 R + \frac{\lambda_1}{2}\sigma^2\right) \phi &= 0,
\label{eq: phi}
\\
\left(\Box + m_2^2 + \xi_2 R + \frac{\lambda_1}{2}\phi^2\right) \sigma + \frac{\lambda_2}{6}\sigma^3&= 0.
\label{eq: sigma}
\end{align}

Additionally, we can compute the classical stress-energy (SE) tensor. 
Since we have written the action on a curved spacetime, we can compute it as customarily done through
\begin{equation}
T_{\mu\nu}:=\frac{2}{\sqrt{g}}\frac{\delta S}{\delta g^{\mu \nu}}= T_{\mu \nu}^{(\sigma)} + T_{\mu \nu}^{(\phi)}  ,
\end{equation}
performing a split that will be useful in the following discussion:
\begin{align}
\begin{split}
T_{\mu \nu}^{(\sigma)}: & = ( 1 - 2 \xi_2) \sigma_{,\mu} \sigma_{,\nu} + (2\xi_2-\tfrac{1}{2}) \eta_{\mu \nu} \sigma_{,\rho}\sigma^{,\rho} - 2 \xi_2 \sigma \sigma_{,\mu \nu}  + 2 \xi_2 \eta_{\mu \nu} \sigma \Box \sigma  + \frac{\eta_{\mu \nu}}{2}\Big(m_2^2 + \tfrac{\lambda_2}{12} \sigma^2\Big)\sigma^2\, ,
\end{split}
\label{eq: Tuv2}
\\
\begin{split}
T_{\mu \nu}^{(\phi)} :& = ( 1 - 2 \xi_1) \phi_{,\mu} \phi_{,\nu} + (2\xi_1-\tfrac{1}{2}) \eta_{\mu \nu} \phi_{,\rho}\phi^{,\rho} - 2 \xi_1 \phi \phi_{,\mu \nu} 
+ 2 \xi_1 \eta_{\mu \nu} \phi \Box \phi + \frac{\eta_{\mu \nu}}{2}\Big(m_1^2 + \tfrac{\lambda_1}{2} \sigma^2\Big)\phi^2\, .
\end{split}
\label{eq: Tuvphi}
\end{align}
After performing the variation,   we 
have set $g_{\mu\nu}=\eta_{\mu\nu}$, since we will not be interested in discussing the interaction with a curved spacetime;
we will work in flat spacetime throughout the rest of the paper.

Two remarks are in oder. First of all, $T_{\mu \nu}^{(\phi)}$
corresponds to the {\se} tensor of a free field with variable mass $M^2=m_1^2 + \frac{\lambda_1}{2} \sigma^2$, 
in agreement with the picture that we have described before.
Secondly,
the full {\se} tensor $T_{\mu\nu}$ is of course conserved classically, while one can easily check 
that\footnote{To simplify the notation, we will adopt the notation $\partial_\nu\sigma^2:=\partial_\nu(\sigma^2)$. }
\begin{equation}\label{eq:class_cons}
\partial^{\mu} T_{\mu\nu}^{(\phi)}= \frac{\lambda_1}{4}\partial_\nu\sigma^2\, \phi^2\, .
\end{equation}

We now consider the semiclassical version of the theory, in which the field $\phi$ is of
quantum nature while $\sigma$ is treated classically. 
Then, the classical expression (\ref{eq: phi}) is promoted to the Heisenberg equation
associated to the quantum operator $\phi$.
On the other side, the evolution equation
for the background field is obtained by taking the vacuum expectation value (VEV) of the classical Eq. (\ref{eq: sigma}),
\begin{equation}
\left(\Box + m_2^2  + \frac{\lambda_1}{2}\vev{\phi^2}\right) \sigma + \frac{\lambda_2}{6}\sigma^3 = 0.
\label{eq: sigma2}
\end{equation}
Additionally, the {\se} tensor of the full semiclassical system is 
\begin{equation}
\vev{T_{\mu\nu}}= T_{\mu \nu}^{(\sigma)} + \vev{ T_{\mu \nu}^{(\phi)}},
\label{eq:Tmunusemi}
\end{equation}
given that  $T_{\mu \nu}^{(\phi)}$ was defined so that it contains all the terms involving the quantum field $\phi$.
Thus, the main objects to analyze the vacuum fluctuations are $\vev{\phi^2}$ and $\vev{T_{\mu \nu}^{(\phi)}}$, 
the latter being relevant to consider Casimir forces and self-energies.
Both of them are divergent 
quantities; as we will see in the following section, the classical action for the field $\sigma$ is needed to absorb the divergences into the bare constants
of the theory during the renormalization process, 
after which we obtain a finite and unique expression for the {\se} tensor (up to finite local terms).
Additionally, we will show in Sec. \ref{sec:conservation_law} that, using the usual prescription, 
Eq. \eqref{eq:class_cons} is valid at the quantum level when the classical quantities are replaced by the corresponding VEVs.


\section{Renormalization and conservation law}\label{sec:ren}

The theory of  quantum fields in curved spacetimes can be renormalized using a precise covariant procedure \cite{Birrel:1982,Parker:2009}.
As was shown in \cite{Paz:1988mt,Mazzitelli:2011st}, the case of a quantum field with a variable mass can be treated in an analogous way;
we will briefly review it in the following.

As customarily in theories with four spacetime dimensions, we can define the renormalized quantities as
\begin{equation}
\begin{split}
\ren{\phi^2}  :& =  \vev{\phi^2}- \ad[\phi^2][2],
\\
\ren{T_{\mu \nu}^{(\phi)}}  :&=  \vev{T_{\mu \nu}^{(\phi)}} - \ad[T_{\mu \nu}^{(\phi)}][4],
\end{split}
\label{eq: renormalizacion}
\end{equation}
where the VEVs $\ad[T_{\mu \nu}^{ (\phi)}][4]$ and $\ad[\phi^2][2]$ 
are constructed using the Schwinger-DeWitt expansion (SWDE) up to fourth and second adiabatic order respectively.
Notice that the counting of the adiabatic order includes not only the number of derivatives, but also the mass dimensions; 
for example, 
a term with $j$ derivatives of $\sigma^2$ is of adiabatic order $j+2$ \cite{Paz:1988mt}. 
After the subtraction in Eq. \eqref{eq: renormalizacion}, 
the divergences in the adiabatic VEVs are to be absorbed into the bare constants of the theory, 
so that we end up with finite renormalized constants and VEVs.

As said above, the adiabatic contributions involve the computation of the SWDE. 
For the Feynman propagator ($G_F$) of a scalar field with mass $m$ one obtains \cite{DeWitt:1965} 
\begin{equation}
\begin{split}
  G_F^{SD}(x,x') 
& = \int\limits_0^{\infty} \frac{\dx[s]}{(4 \pi \mathi s)^{\dime/2}} e^{-\frac{\mathi \sigma_S(x,x')}{2s}-\mathi (m^2-\mathi \epsilon)s} \sum_{j \geq 0}(\mathi s)^j \Omega_j(x,x'),
\end{split}
\label{eq:SDW2}
\end{equation}
where $\dime$ is the number of spacetime dimensions and $\sigma_S(x,x')$ is 
Synge's world function, that in flat space is just $\sigma_S(x,x') = (x-x')^2/2$.
The functions  $\Omega_j(x,x')$ are defined by a set of recursive equations that follow from imposing
the equation for the propagator, i.e.
\begin{equation}
\label{ec:prop_def_0}
    \left[ \Box +m_1^2+ \frac{\lambda }{2} \sigma^2(x)-\mathi \epsilon \right] G_F(x,x') = - \delta^4(x-x')\, .
\end{equation}
Their general form is well-known; denoting with square brackets
the coincidence limit of these functions and their derivatives
it can be shown that, for the action in Eq. \eqref{eq: accion de materia} in flat spacetime,
the first functions read\footnote{\label{foot:derivatives}It should be understood that $\Omega_{1 \, ,\mu \nu}=\frac{\partial}{\partial x^\mu}\frac{\partial}{\partial x^\mu} \Omega_1(x,x')$.
Notice that Ref. \cite{Vinas:2010ix} works with an Euclidean signature.} \cite{Vinas:2010ix, Vassilevich:2003xt} 
\begin{eqnarray}
&& \Omega_0(x,x')=1,
\quad [\Omega_1]=-\tfrac{\lambda_1}{2}\sigma^2,
\quad [\Omega_{1\, ,\mu\nu}]=-\tfrac{\lambda_1}{6}\sigma^2_{,\mu\nu},
\quad [\Omega_2]=\tfrac{\lambda_1}{12}\Box\sigma^2 +\tfrac{\lambda_1^2}{8}
\sigma^4\,.
\end{eqnarray}
This expansion can  be  modified by including the full variable mass $M^2$ 
into the exponent of the SDWE in Eq. \eqref{eq:SDW2} (see \cite{Paz:1988mt} and more recently \cite{Ferreiro:2020zyl,Ferreiro:2020uno}).
With this modification, we will have an expansion analogue to expression \eqref{eq:SDW2} 
involving new functions $\tilde \Omega_j(x,x')$; the latter do not contain powers of $M^2$, 
but only powers of its derivatives.  
Although this expansion could be used in principle\footnote{One should appropiately modify the discussion in Sec. \ref{sec:conservation_law}.}, 
it does not provide additional help in the following computations and will not be followed here.

Coming back to the computation of the renormalized quantities, 
the adiabatic VEVs can be computed by recasting all the expressions in terms of the imaginary part of Feynman's Green function,
which satisfies
\begin{equation}
\mbox{Im}\left(G_F(x,x')\right)=-\tfrac{1}{2}\langle\{\phi(x),
\phi(x´)\}\rangle\, .
\end{equation}
A direct computation shows that the explicit expressions are (see \cite{Paz:1988mt} and footnote \ref{foot:derivatives} above)
\begin{align}\label{eq: phi2ImGf}
\vev{\phi^2}=& - \mbox{Im} \left[ {G_F}\right],
\\
\begin{split}
\vev{T_{\mu \nu}^{(\phi)}}   = & -  \, \mbox{Im}\left\{ - [G_{F \, ,\mu \nu}] +  (\tfrac{1}{2} - \xi_1) [G_F]_{,\mu \nu} 
+ (\xi_1- \tfrac{1}{4})\eta_{\mu \nu} \Box [G_F] \right\}\, .
\end{split}
\label{eq: TuvG}
\end{align}
In these expressions one can replace the SDWE \eqref{eq:SDW2} for the propagator 
and obtain the adiabatic expansion of the desired quantities up to the appropriate order. 
 
In particular, the coincidence limit of the two-point function 
for a field with variable mass as in Eq. \eqref{eq: accion de materia} is therefore given by 
\begin{equation}
\ad[\phi^2][2] = \frac{1}{(4 \pi)^{\dime/2}} 
\left[\frac{m_1}{\mu}\right]^{\dime-4}\left\{m_1^2\, \Gamma\left(1-\tfrac{\dime}{2}\right)-\frac{\lambda_1}{2}\sigma^2\, \Gamma\left(2-\tfrac{\dime}{2}\right)\right\},
\label{eq: phi2 ad2}
\end{equation}
where $\mu$ is an arbitrary scale with dimensions of mass that is introduced in the renormalization process.
Both terms diverge as $\dime \to 4$ and their subtraction will be enough to obtain a finite result in \eqref{eq: renormalizacion}. 
The fact that the divergences can be absorbed into the bare constants of the theory can be seen 
by inserting  expression \eqref{eq: phi2 ad2} into the semiclassical equation for $\sigma$, i.e. Eq. into \eqref{eq: sigma2}. Indeed, writing 
\begin{equation}
\begin{split}
m_2^2 & =:m_{2R}^2+\delta m_2^2,
\\
\lambda_2 & =: \lambda_{2R}+\delta \lambda_2,
\end{split}
\label{eq: reescritura}
\end{equation}
we obtain the counterterms 
\begin{equation}
\begin{split}
\delta m_2^2 & = -\frac{\lambda_1 m_1^2}{16 \pi^2 (\dime-4)}+ \Delta m_1^2,
\\
\delta \lambda_2 & =  - \frac{3 \lambda_1^2}{16 \pi^2 (\dime-4)} + \Delta \lambda_2
\, ,
\label{eq: redefiniciones1}
\end{split}
\end{equation}
where $\Delta m_2^2$ and $\Delta \lambda_2$ are finite contributions that relate different renormalization schemes (they vanish in the minimal 
subtraction scheme). 

We now consider the evaluation of the {\se} tensor in our semiclassical theory.  The expression for its VEV up to fourth adiabatic order reads  
\begin{align}
\begin{split}
\ad[T_{\mu \nu}^{(\phi)}][4]&=
\frac{1}{(4\pi)^{\dime/2}}\left[\frac{m_1}{\mu}\right]^{\dime-4}\bigg\{
-\frac{\eta_{\mu\nu}}{2}m_1^4\, \Gamma\left(-\tfrac{\dime}{2}\right)+\frac{\lambda_1}{4}m_1^2\sigma^2\eta_{\mu\nu} \, \Gamma\left(1-\tfrac{\dime}{2}\right)
\\
&\hu\hu+ \Gamma\left(2-\tfrac{\dime}{2}\right)\left[-\frac{\lambda_1^2}{16}\eta_{\mu\nu}\sigma^4+\frac{\lambda_1}{2}\left(\xi_1-\frac{1}{6}\right)\left(\sigma^2_{,\mu\nu}-\eta_{\mu\nu}\Box\sigma^2\right)\right]\bigg\}\, .
\label{eq: Tuv div}
\end{split}
\end{align}
Comparing Eq. (\ref{eq: Tuv div}) with Eq. \eqref{eq: Tuv2} one can show that the $\sigma$-dependent divergences can be absorbed using the same counterterms given in Eq. (\ref{eq: redefiniciones1}) and
including a counterterm for $\xi_2$, the latter needed to absorb the divergence proportional to $(\xi_1-1/6)$. 
The term independent of $\sigma^2$ will just renormalize the cosmological constant (or a bare constant in the classical potential for the background field), and will play no role in our considerations.
All these terms depend on the arbitrary scale $\mu$ that has been introduced in the renormalization process; the  arbitrariness is resolved by using experimental data to fix the involved couplings.
Therefore, we have a precise
procedure for defining the renormalized {\se} tensor for the quantum field $\phi$ in an inhomogenoeus background $\sigma$. 

Since in the following sections we will deal with massless fields, let us recall that then one can simply trade $m_1^{\dime-4}\to \mu_2^{\dime-4}$ in Eqs.
\eqref{eq: Tuv div} and \eqref{eq: phi2 ad2}, setting the other powers of $m_1$ to zero. 
The new scale $\mu_2$ is arbitrary but appears only in quotients with $\mu$; for convenience, we can set it to $\mu_2\equiv e^{-1}\mu$.

Before concluding this section one last remark is in order. Our starting action in Eq. \eqref{eq: accion de materia}, 
belongs to a theory on curved spacetime.  
This choice was motivated in part to
emphazise that the problem of quantum fields in inhomogeneous backgrounds can be addressed 
using well-known techniques of quantum fields on curved spaces. 
However, this point is not crucial from a computational point of view.
An alternative route is to start with the theory in Minkowski spacetime and compute the {\se} tensor using Noether's
theorem; afterwards one may add the terms proportional to $\xi_1$ and $\xi_2$
in Eqs. \eqref{eq: Tuv2} and \eqref{eq: Tuvphi} using the fact that Noether's theorem does not constrain them. 
In any case, note that while it is not necessary to add the terms proportional to 
$\xi_1$ in the {\se} tensor of the quantum field $\phi$, 
the introduction of the classical terms proportional to 
$\xi_2$ is essential to renormalize the theory, even if
$\xi_{1}=0$.

\subsection{Semiclassical conservation law}\label{sec:conservation_law}

It is well-known that the renormalization procedure may induce anomalies in the quantum theory, 
which may be caused by the regularization and/or the corresponding subtractions. 
Typical examples are the non-conservation of the chiral current for massless fermions
in the presence of background gauge fields \cite{Adler:1969gk} and the trace anomaly for conformal fields in curved spaces, 
first discovered in  \cite{Capper:1973mv} and lately revisited in relation with Weyl fermions \cite{Abdallah:2021eii,Bonora:2017gzz}. 
We will now show that
the conservation law  in Eq. \eqref{eq:class_cons} 
remains valid after the quantization of $\phi$, 
if we replace the classical quantities with the corresponding renormalized VEVs given by expression \eqref{eq: renormalizacion}:
\begin{equation}\label{eq:cons sem}
\partial^{\mu} \ren{T_{\mu\nu}^{(\phi)}}= \frac{\lambda_1}{4}\partial_\nu\sigma^2\, \ren{\phi^2}.
\end{equation}
To see this, we rewrite the above equation as 
\begin{equation}
\partial^{\mu}\Big(\vev{T_{\mu\nu}^{(\phi)}}-\ad[T_{\mu\nu}^{(\phi)}][4]\Big)= \frac{\lambda_1}{4}\partial_\nu\sigma^2\,\Big(\vev{\phi^2}- \ad[\phi^2][2]\Big)\, ,
\end{equation}
where all calculations are performed in $\dime$ dimensions. 
As dimensional reguarization is covariant, one expect the regularized mean values $\vev{T_{\mu\nu}^{(\phi)}}$
and $\vev{\phi^2}$ to satisfy the conservation law. 
Indeed, from Eqs. \eqref{eq: phi2ImGf}
and \eqref{eq: TuvG} one can check this explicitly  using the expression for the propagator, which is of course valid 
in $\dime$ dimensions. Moreover, computing 
the derivative of $\ad[T_{\mu\nu}^{(\phi)}][4]$
in Eq. \eqref{eq: Tuv div} we straightforwardly obtain
\begin{equation}
\partial^{\mu} \ad[T_{\mu\nu}^{(\phi)}][4]= \frac{\lambda_1}{4}\partial_\nu\sigma^2\, \ad[\phi^2][2]\, ,
\end{equation}
so neither the regularization nor the subtraction breaks the conservation law at the quantum level for $\phi$. Therefore,
Eq. \eqref{eq:cons sem} is valid. This will be crucial in the discussion of
the principle of virtual work in Sec. \ref{sec:PVW}.


\section{The perturbative approach}\label{sec:perturbative}

We will now obtain explicit expressions for the renormalized VEVs
$\ren{T_{\mu\nu}^{(\phi)}}$ and $\ren{\phi^2}$, 
using a perturbative expansion in powers of $\lambda_1$. 
We will start studying Feynman's propagator, since  those VEVs can be obtained from it as shown in Eqs. \eqref{eq: phi2ImGf} and 
\eqref{eq: TuvG}. For simplicity we will consider the massless case ($m_1^2\equiv0$) and replace $\lambda_1\to\lambda$, 
so that using the customary ``$\mathi \epsilon$'' prescription  Feynman's propagator satisfies
\begin{equation}
\label{ec:prop_def}
    \left[ \Box + \frac{\lambda }{2} \sigma^2(x)-\mathi \epsilon \right] G_F(x,x') = - \delta^4(x-x')\,.
\end{equation}
Solving this equation perturbatively in $\lambda$ we obtain
\begin{equation}
 G_F(x,x') =  G_F^{(0)}(x,x')+ G_F^{(1)}(x,x')+ \cdots,
 \end{equation}
defining $G_F^{(0)}(x,x')$ as the usual free propagator and the contribution of order $\lambda^n$ as
\begin{equation}\label{ec:prop_orden_n_conf}
    G_F^{(n)}(x,x') := \frac{\lambda^n}{2^n} \int \measure{x}\, \sigma^2(x_1) \cdots \sigma^2(x_n) G_F^{(0)}(x,x_n) \cdots G_F^{(0)}(x_1,x').
\end{equation}
This is a notation that we will employ frequently in the following: the order $n$ contribution in $\lambda$ of a given quantity will be denoted by adding a superscript $(n)$. 
Coming back to \eqref{ec:prop_orden_n_conf}, we can recast it by expressing every free propagator in momentum space:
\begin{equation}
\begin{split}
\label{ec:prop_orden_n}
    G_F^{(n)}(x,x') &= \frac{\lambda^n}{2^n (2 \pi)^{4n+4}}  \int \measure{x}\measure{q} \,\sigma^2(x_1) \cdots  \sigma^2(x_n)   \\
    & \qquad \times e^{-\mathi q_n\cdot (x_n-x_{n-1})} \cdots e^{-\mathi  q_1\cdot (x_1-x')}  \int  \frac{\dxd[s][\dime]}{\mu^{\dime-4}} \; \frac{e^{-is(x-x')}}{s^2 (s+q_1)^2 \cdots (s+q_n)^2}.
\end{split}
\end{equation}
Notice that we have implemented dimensional regularization only in the internal momentum $s$, introducing as usual an arbitrary scale $\mu$ with dimensions of mass; to shorten the notation, we have omitted the $\mathi \epsilon$ terms in the propagators.
An immediate consequence  of this result is that
\begin{equation}
\begin{split}
\label{ec:phi_orden_n}
     \vev{\phi^{2}}^{(n)}  &= -\frac{\lambda^n}{2^n (2 \pi)^{4n+4}}   \operatorname{Im} \bigg[ \int \measure{x}\measure{q}\, \sigma^2(x_1) \cdots \sigma^2(x_n)  \\
    & \hspace{5.5cm}\times e^{-\mathi q_n\cdot (x_n-x_{n-1})} \cdots e^{-\mathi q_1\cdot (x_1-x)}   \mathcal{I}(q_1,\cdots,q_n)\bigg],
\end{split}
\end{equation}
where we have introduced the tensorial integrals\footnote{The integral $\mathcal{I}$ without indices should be understood with a factor $1$ in the integrand's numerator of \eqref{eq:I_integral}.}
\begin{equation}\label{eq:I_integral}
 \mathcal{I}^{\mu_1,\mu_2\cdots}(q_1,\cdots,q_n):= \int \frac{\dxd[s][\dime]}{\mu^{\dime-4}}  \frac{ s^{\mu_1}s^{\mu_2}\cdots }{s^2(s+q_1)^2(s+q_2)^2\cdots(s+q_n)^2}\, .
\end{equation}
The computation of these integrals can be done in various ways, the most famous one being probably the Veltman-Passarino reduction method \cite{Passarino:1978jh} (see also \cite{tHooft:1978jhc, smirnov, Abdallah:2021eii}). 

An analogous expansion for the $\phi$ contribution to the {\se} tensor can be obtained.
Inserting the $n$-th order expression for the propagator 
into Eq. \eqref{eq: TuvG} and dropping the $(\phi)$ superscript one can find 
\begin{equation}
\begin{split}
\label{ec:tmunu_orden_n}
    &\vev{T_{\mu \nu}}^{(n)}= \frac{\lambda }{4} \eta_{\mu \nu} \sigma^2(x) \vev{\phi^{2}}^{(n-1)}
    -\frac{\lambda^n}{2^n (2 \pi)^{4n+4}}   \operatorname{Im}\Bigg\{ \int \dxd[x_1][4]\cdots \dxd[q_1][4]\cdots \,\sigma^2(x_1) \cdots \sigma^2(x_n)
    \\
    & e^{-\mathi q_n\cdot (x_n-x_{n-1})} \cdots e^{-\mathi q_1\cdot (x_1-x)}
    \left[ \left(\mathcal{I}_{\rho\sigma } + \mathcal{I}_{\rho} q_{1\sigma}\right) \left(\eta^\rho_\mu \eta^\sigma_\nu - \frac{1}{2} \eta^{\rho \sigma} \eta_{\mu \nu} \right)
    -2 \xi  \mathcal{I}_{\rho }  q_{1 \sigma} \left(\eta^\rho_\mu \eta^\sigma_\nu - \eta^{\rho \sigma} \eta_{\mu \nu} \right) \right] \Bigg\} \, ,
\end{split}
\end{equation}
where it will be understood that the arguments in the $\mathcal{I}$ tensorial integrals, when missing, are all the involved momenta $q_1,\,\cdots,q_n$.

From now on we will assume that the background field is time-independent. In that case, integrating over all space we find an expression for the total
vacuum energy, $E_{}$, that reads
\begin{align}\label{eq:energy_n_time_independent}
\begin{split}
    &E^{(n)}_{}
    = 
    -\frac{\lambda^n}{2^n (2 \pi)^{3n+1}}   \operatorname{Im}\Bigg[ \int  \dxd[\alls[x]_1][3]\cdots \dxd[\alls[q]_2][3]\cdots \,\sigma^2(\vec{x}_1) \cdots \sigma^2(\vec{x}_n)
    \\
    &\times e^{-\mathi\alls[q]_n\cdot (\alls[x]_n-\alls[x]_{n-1})} \cdots e^{-\mathi\alls[q]_2\cdot (\alls[x]_2-\alls[x]_1)}
     \left(\mathcal{I}_{00 } - \frac{1}{2} \mathcal{I}^{\rho}{}_{\rho} \right)\Bigg]_{q_1^{\mu}=0} +\frac{\lambda }{4}  \int \dxd[\alls[x]][3]\,  \sigma^2(\vec{x})  \vev{\phi^{2}}^{(n-1)}(\vec{x})
      \, .
\end{split}
\end{align}
This expression can be further simplified. 
First of all, the term involving $\mathcal{I}^{\rho}{}_{\rho} $ cancels with the one proportional to $\vev{\phi^{2}}^{(n-1)}$.
Second, $\mathcal{I}_{00 }$ can be recast integrating by parts in the zeroth component of the internal momentum $s$, 
using the symmetry of the integrand in the variables $q_i$
and rewriting the result in terms of $\vev{\phi^{2}}^{(n-1)}$.
This leads to the following  master formula for the time-independent case:
\begin{equation}\label{ec:energy_orden_n_fin}
\begin{split}
    E^{(n)}_{} = & \frac{\lambda}{4 n} \int \dxd[\alls[x]][3]\,\sigma^2(\vec{x}) \,\vev{\phi^2}^{(n-1)}(\vec{x}).
\end{split}
\end{equation}
It is important to notice that, with our renormalization prescription, Eq. \eqref{ec:energy_orden_n_fin} remains valid when replacing the regularized quantities by the renormalized ones.

In the next sections we will derive explicit expressions for all the relevant physical quantities at first and second order in $\lambda$, 
along with some illustrative examples. 
Before doing this, we would like to stress some general properties of the preceding results. 

In Section \ref{sec:ren} we have discussed
the divergences' structure  of the VEVs
$\vev{\phi^2}$ and $\vev{T_{\mu\nu}}$. 
The ones that should be renormalized are at most quadratic in the coupling constant
$\lambda$ and therefore we will be able to reproduce them in a second-order perturbative approach. 
After subtracting the appropiate adiabatic expansions, 
the renormalized VEVs will be determined up to local terms whose dependence on $\sigma$ is that of the counterterms. 
Since they are local, they are not relevant in the computation of Casimir forces between different bodies;
in other words, Casimir forces will have no undeterminacy. However, if one were interested in self-energies, then one should use experiments to fix the otherwise free parameter $\mu$.

One subtle point is that there could be additional divergences. 
First,  they could be generated by discontinuities
in the background field $\sigma^2$ or its derivatives. 
These are the scalar counterparts of those arising near a perfect conductor, 
which depend on the local geometry of the
surface \cite{Deutsch:1978sc}.
In these situations, one should be careful to give the right interpretation of the conservation equation \eqref{eq:cons sem}. 
We will describe this kind of divergences in Sec. \ref{sec:perturbative_examples}
below.

We would also like to point out that, due to the fact that we are using massless propagators,
one could encounter infrared divergences at higher orders in $\lambda$. 
In order to avoid these divergences, 
one could consider  massive propagators. 
That will be the case if the field $\phi$ is massive. Alternatively, for a massless field, one can perform the perturbative expansion around the average of $\sigma^2$ over all space
($\bar\sigma^2$). If the latter is nonvanishing, one can write the equation for the propagator
as
\begin{equation}
\label{ec:prop_def_mass}
    \left[ \Box + \frac{\lambda }{2} \bar \sigma^2 -\mathi\epsilon + \frac{\lambda }{2} (\sigma^2(x)-\bar\sigma^2) \right] G_F(x,x') = - \delta^4(x-x')\, 
\end{equation}
and perform the expansion with a free 
propagator of mass $\bar m^2=\tfrac{\lambda }{2} \bar \sigma^2$. This 
corresponds to a resumation of the perturbative results, that
will show a non-analytic dependence with $\bar m^2$. In both cases, the corresponding perturbative expressions can be obtained just by replacing in Eqs. \eqref{ec:prop_orden_n_conf} to \eqref{ec:energy_orden_n_fin} the free massless propagators by massive ones.

Finally, we would like to point out  that the perturbative approach should be modified when considering a time-dependent background field. 
Indeed, the solution to Eq.
\eqref{ec:prop_def} is the matrix element
\begin{equation}
G_F(x,x')=\mathi\frac{\langle 0_{\rm IN}\vert T\big(\phi(x)\phi(x') \big)\vert 0_{\rm OUT}\rangle}{\langle 0_{\rm IN}\vert 0_{\rm OUT}\rangle},
\end{equation}
which involves the initial and final vacuum states, not the mean value $\langle 0_{\rm IN}\vert T\big(\phi(x)\phi(x') \big)\vert 0_{\rm IN}\rangle$. 
The same remark applies to the other VEVs in this section. This situation can be amended following
a procedure inspired in the Schwinger-Keldysh formalism \cite{Calzetta:2008iqa},
by computing perturbatively the generalized Green function
\begin{equation}
G_{\cal C}(x,x´) = \mathi\, \langle 0_{\rm IN}\vert T_{\cal C}\big(\phi(x)\phi(x') \big)\vert 0_{\rm IN}\rangle\, ,
\end{equation}
where $T_{\cal C}$ is the temporal ordering along a closed temporal contour $\cal C$. This is beyond the scope of the present paper.


\section{The principle of virtual work}\label{sec:PVW}

\begin{figure}[h!]
    \centering
    \includegraphics[width=.5\columnwidth]{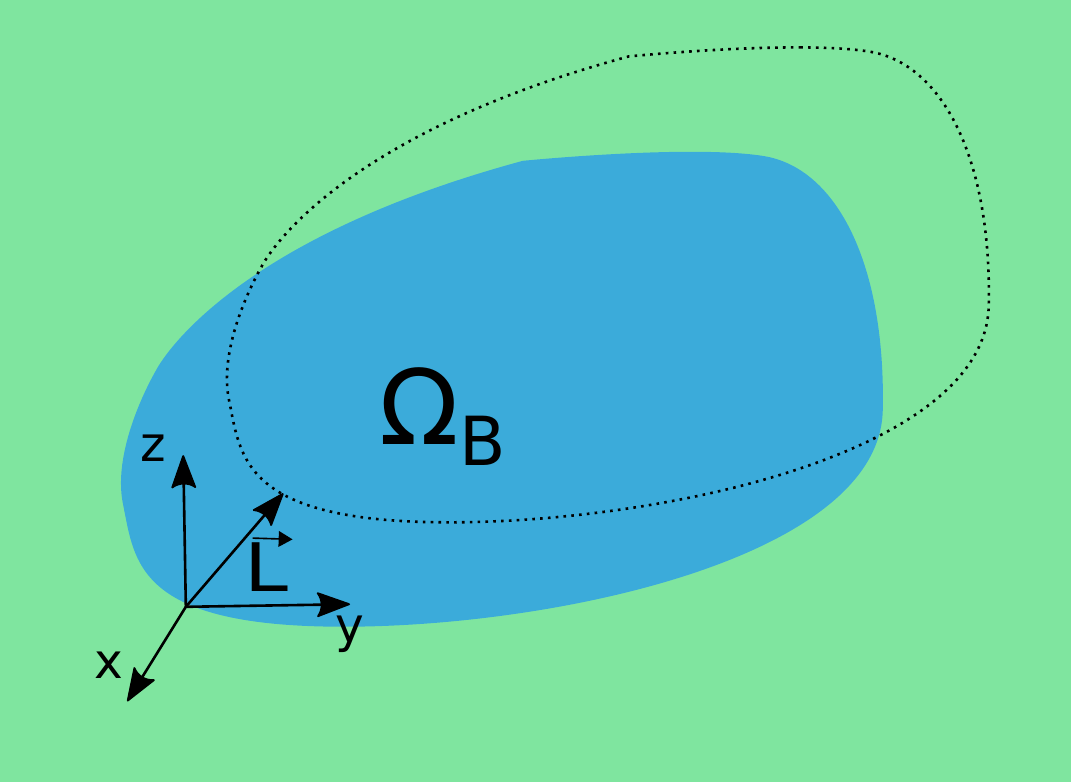}
    \caption{The body B, that initially occupies the volume $\Omega_B$, is virtually displaced by the vector  $\vec{L}$.}
    \label{fig:pvw_cuerpo_inmerso}
\end{figure}

Before we apply our  formulas in Sec. \ref{sec:perturbative_examples} to some particular configurations, we will  provide an explicit  proof of the validity of the 
PVW in this model. To do that, we consider the situation is illustrated in Fig. \ref{fig:pvw_cuerpo_inmerso}, in which a body is immersed in an inhomogeneous media. 
Then we compare the variation of the energy under an infinitesimal displacement of the body $B$ and the integral of the normal component of the {\se} tensor over the surface of the same body.

Let us denote by $\Omega_B$ the volume occupied by the body
in the initial position.
The body $B$ is caractherized by a field $\sigma^2_B(\vec{x})$ 
while the sorrounding media corresponds to $\sigma^2_M(\vec{x})$.
Introducing the characteristic function
\begin{equation}
    \scalebox{1.2}{$\chi$}_{\Omega} (\vec{x}) = \begin{cases} 
      1 & \vec{x} \in \Omega \\
      0 & \vec{x} \notin \Omega
    \end{cases},
\end{equation}
it is clear that after a translation by a vector $\vec{L}$ the background field becomes
\begin{align}
\sigma^2_{\vec{L}}(\vec{x}) &= \sigma_B^2(\vec{x} - \vec{L}) \scalebox{1.2}{$\chi$}_{\Omega_B} (\vec{x} - \vec{L}) + \sigma_M^2(\vec{x})[1 - \scalebox{1.2}{$\chi$}_{\Omega_B} (\vec{x} - \vec{L})],
\end{align}
which is different from $\sigma^2(\vec{x-\vec{L}})$. 
In such affirmation we are assuming that the effects of one media on the other, if they exist, can be neglected in the evaluation of response functions.
We are also supposing that the function $\sigma^2_M$ is defined over all space, independently of the presence of the body $B$.

We now consider the gradient $\nabla$ of  the background field with respect to  $\vec{x}$,
\begin{equation}
\begin{split}
    \nabla \sigma^2(\vec{x}) &= \nabla \left[ \sigma_B^2(\vec{x} ) \scalebox{1.2}{$\chi$}_{\Omega_B} (\vec{x}  ) + \sigma_M^2(\vec{x})\left( 1 - \scalebox{1.2}{$\chi$}_{\Omega_B} (\vec{x} ) \right)\right].
\end{split}
\end{equation}
Computing the gradient of $\sigma^2_{\vec{L}}(\vec{x})$ with respect to $\vec{L}$ at zero displacement,
\begin{equation}
\begin{split}
    \frac{\partial}{\partial \vec{L}} \sigma^2_{\vec{L}}(\vec{x}) \bigg\vert_{\vec{L}=0}&= \frac{\partial}{\partial \vec{L}} \left[ \sigma_B^2(\vec{x} - \vec{L}) \scalebox{1.2}{$\chi$}_{\Omega_B} (\vec{x} - \vec{L}) \right] - \sigma_M^2(\vec{x}) \frac{\partial}{\partial \vec{L}}  \scalebox{1.2}{$\chi$}_{\Omega_B} (\vec{x} - \vec{L})\bigg\vert_{\vec{L}=0}
    \\
    &=-\nabla \left[ \sigma_B^2(\vec{x} - \vec{L}) \scalebox{1.2}{$\chi$}_{\Omega_B} (\vec{x} - \vec{L}) \right] + \sigma_M^2(\vec{x}) \nabla \scalebox{1.2}{$\chi$}_{\Omega_B} (\vec{x} - \vec{L})\Big\vert_{\vec{L}=0}\, ,
\end{split}
\end{equation}
it is immediate to see that $\frac{\partial}{\partial \vec{L}} \sigma_{\vec{L}}^2(\vec{x})\vert_{\vec{L}=0}$ 
is non vanishing only in the region
 $\Omega_{B}$ (including the boundary), and in that
 region 
 \begin{align}\label{eq:pvw_divergences}
\frac{\partial}{\partial \vec{L}} \sigma_{\vec{L}}^2(\vec{x})\bigg\vert_{\vec{L}=0} = - \nabla \sigma^2(\vec{x}),\quad \alls[x]\in \Omega_B.  
 \end{align}

In the time-independent situation, one can compute the energy after a virtual displacement of the body by replacing $\sigma^2$ with $\sigma_{\vec{L}}^2$ in Eq. \eqref{eq:energy_n_time_independent}. Afterwards, taking  the derivative of the energy with respect to $\vec{L}$
and using the symmetry of the integrand we get
\begin{equation}
\begin{split}
    - \frac{\partial E^{(n)}_{}}{\partial \vec{L}}  \bigg\vert_{\vec{L}=0}&=  \frac{1}{2} \frac{\lambda^n}{2^n (2 \pi)^{3n+1}} \operatorname{Im} \Bigg[  \int \dxd[\alls[x]_1][3] \cdots \dxd[\alls[q]_2][3] \cdots  \frac{\partial \sigma^2_{\vec{L}}}{\partial \vec{L}}(\vec{x}_1) \, \sigma_{\vec{L}}^2(\vec{x}_2) \cdots  \\
    &\hspace{4cm}\times e^{i \vec{q}_n \cdot (\vec{x}_n-\vec{x}_{n-1})} \cdots e^{i \vec{q}_2\cdot (\vec{x}_2-\vec{x}_1)}\, \mathcal{I}(q_2,\cdots)  \Bigg]_{q_k^0 = 0,\,{\vec{L}}=0}.
\end{split}
\end{equation}
As previously done  with the energy, we can rewrite this expression in terms of $ \vev{\phi^2 (\vec{x})}^{(n-1)}$ as follows:
\begin{equation}
    - \frac{\partial E^{(n)}_{}}{\partial \vec{L}} \bigg\vert_{\vec{L}=0} = - \frac{\lambda}{4} \int \dxd[\alls[x]][3]\, \vev{\phi^2 (\vec{x})}^{(n-1)}\, 
    \frac{\partial \sigma^2_{\vec{L}}}{\partial \vec{L}}(\vec{x})\bigg\vert_{\vec{L}=0}.
\end{equation}
Recalling from Eq. \eqref{eq:pvw_divergences} that $\frac{\partial \sigma^2_{\vec{L}}}{\partial  \vec{L}}(\vec{x}) \big\vert_{\vec{L}=0}$ is different from
zero only for $\vec{x}\in \Omega_{B}$, we may replace 
the derivatives with respect to the displacement by minus the gradient and obtain
\begin{equation}
\begin{split}
    - \frac{\partial E^{(n)}_{}}{\partial \vec{L}} \bigg\vert_{\vec{L}=0} &=  \frac{\lambda}{4} \int_{\Omega_{B}} \dxd[\vec{x}][3] \, \nabla \sigma^2(\vec{x}) \,\vev{\phi^2 (\vec{x})}^{(n-1)},
\end{split}
\end{equation}
{where the integral over $\Omega_B$ includes possible surface-localized  contributions.}
Comparing this expression with the conservation law \eqref{eq:cons sem} of the SE tensor 
for a static configuration\footnote{Latin indeces are used for spatial coordinates},
\begin{equation}
 \partial_i\vev{T^{i}{}_{j} (\vec{x})}^{(n)} = \frac{\lambda}{4} \partial_j\sigma^2(\vec{x})\, \vev{\phi^2 (\vec{x})}^{(n-1)}, 
 \end{equation}
we have therefore 
\begin{equation}
\begin{split}
    - \frac{\partial E^{(n)}_{}}{\partial {L}^{j}} \bigg\vert_{\vec{L}=0} &=  \int_{\Omega_{B}} \dxd[\vec{x}][3] \,\partial_i\vev{T^{i}{}_{j}(\vec{x})}^{(n)}.
\end{split}
\end{equation}
{If $\partial_i\vev{T^{i}{}_{j}(\vec{x})}^{(n)}$ is regular enough, one can then prove the PVW by using Gauss' theorem; calling $\dxd[\Sigma_i][2]$ the positive volume 1-form on $\partial\Omega_B$, we obtain
\begin{equation}\label{eq:pvw}
\begin{split}
    - \frac{\partial E^{(n)}_{}}{\partial  {L}^{j}}\bigg\vert_{\vec{L}=0}  &=  \int_{\partial \Omega_{B}} \dxd[\Sigma_i][2] \,  \vev{T^{i}{}_j(\vec{x})}^{(n)}.
\end{split}
\end{equation}
The extension of the proof to the renormalized VEVs of the {\se} tensor can be done by showing that the subtracted adiabatic terms satisfy an equation analogous to \eqref{eq:pvw}.
Notice that if $\partial_i\vev{T^{i}{}_{j}(\vec{x})}^{(n)}$ has surface-localized contributions on $\partial\Omega_B$ then they should be added to the RHS of  Eq.~\eqref{eq:pvw}.}


\section{Examples}\label{sec:perturbative_examples}

\subsection{First-order perturbation theory}\label{sec:order1}

The first-order expressions have been previously  obtained in \cite{Mazzitelli:2011st}. 
The divergent parts can be straightforwardly obtained in our formalism 
by computing the involved scalar integral $\mathcal{I}$;
they agree with those predicted by the adiabatic expansions \eqref{eq: phi2 ad2} and \eqref{eq: Tuv div}.
Furthermore, one can obtain an explicit result for the renormalized quantities:
\begin{align}\label{eq:phi2_order1}
\begin{split}
 \ren{\phi^2}^{(1)} &= \frac{\lambda \pi^2}{2 (2\pi)^8} \operatorname{Re} \Bigg\lbrace\int  \dxd[q_1][4]\, e^{\mathi \,q_1 \cdot  x}\,\tilde \sigma^2(q_1) \, \log\left(-\frac{q_1^2}{\mu^2}-\mathi \epsilon\right)\Bigg\rbrace,
 \end{split}
\\
\label{eq:EM_order1}
 \begin{split}
\ren{T^{\mu\nu}}^{(1)} &= -\frac{\lambda}{48(2\pi)^6} \text{Re} \Bigg\lbrace \int \dxd[q_1][4]\, e^{\mathi\, q_1 \cdot  x} \, \tilde \sigma^2(q_1)\,(q_{1}^{\mu}q_{1}^{\nu}-q_1^2 \eta_{\mu\nu}) \,\log\left( -\frac{q_1^2}{\mu^2}-\mathi \epsilon\right)\Bigg\rbrace
 \\
 &\hu{+\frac{\lambda}{144 (2\pi)^2 }\Big[\partial^{\mu}\partial^{\nu}- \eta^{\mu\nu} \Box \Big] \sigma^2(x)} ,
 \end{split}
\end{align}
where we have made explicit the ``$+\mathi \epsilon$'' prescription and we have defined a Fourier transform in Minkowski space as 
\begin{align}
  \tilde \sigma^2(q):&=\int \dxd[x_{1}][4]\, e^{-\mathi \,q \cdot x_1}   \sigma^2(x_1) .
 \end{align}
Although one could be tempted to cancel the last term in the RHS of Eq. \eqref{eq:EM_order1} by performing a redefinition of the renormalization scale $\mu$, that would imply the introduction of an additional term in other quantities, such as expression \eqref{eq:phi2_order1}. Related to this fact, the choice of $\mu_2$ made in Sec. \ref{sec:ren} is such that there are no local terms in the expression for $\ren{\phi^2}^{(1)}$ (apart from the $\mu$ dependent ones).

The first order approximation for the {\se} tensor satisfies $\partial_\mu\ren{T^{\mu\nu}}^{(1)}=0$, which is consistent with the conservation law 
in Eq. \eqref{eq:cons sem} up to order $\lambda$,
given that $\ren{\phi^2}^{(0)}=0$.
It could ber useful to analyze the eventual gravitational effects of the vacuum fluctuations, when used as a source in the semiclassical Einstein equations \cite{Mazzitelli:2011st}. 
However, the associated vacuum energy $E^{(1)}$ vanishes for static backgrounds and therefore  has no relevance in the computation of non-dynamical Casimir forces.

\subsection{Second-order perturbation theory}\label{sec:order2}
The computation at second order in $\lambda$ is more challenging.
After introducing Feynman parameters we are able to isolate the divergences in the $\mathcal{I}$ integrals 
and perform the corresponding renormalization;
afterwards, we obtain the following results for the  contributions:
\begin{align}
\begin{split}
\label{ec:phi_orden_2}
     \ren{\phi^{2}}^{(2)}  &= -\frac{\lambda^2}{2^2 (2 \pi)^{12}}   \operatorname{Re} \bigg[ \int \dxd[q_{1}][4]\dxd[q_{2}][4]\, \tilde\sigma^2(q_1-q_2)\tilde \sigma^2(q_2) 
     e^{\mathi\, q_1\cdot x}  
    \\
    &\hspace{6.5cm}\times\int_0^1 \int_0^{1-s_1}\dx[s_2]\dx[s_1]  \left(M_E+\mathi \epsilon \right)^{-1} \bigg],
\end{split}
\\
\label{eq:EM_order2}
 \begin{split}
\ren{ T{}^{\mu\nu}}^{(2)}
 &=
  -\frac{\lambda^2}{16(2\pi)^{10}} \text{Re}\Bigg\lbrace \int_0^1\int_0^{1-s_1} \dx[s_2]\dx[s_1] \int \dxd[q_1][4]\dxd[q_2][4]\, \tilde\sigma^2(q_1-q_2)\tilde \sigma^2(q_2) e^{\mathi \,q_1 \cdot x}
 \\
 &
  \hspace{-1cm}\times\Bigg[\frac{\eta^{\mu\nu}}{2 } \log \left(\frac{M_E}{\mu^2}+\mathi \epsilon\right) + \frac{((s_1-1)q_1+s_2q_2)^{\rho}(s_1q_1+s_2q_2)^{\sigma}}{M_E+\mathi0 } 
 \Big(\eta^{\mu}_{\rho}\eta^{\nu}_{\sigma}-\frac{1}{2} \eta^{\mu\nu}\eta_{\rho\sigma}\Big) \Bigg]
 \Bigg\rbrace
  \\
 &\hspace{-1cm}+\frac{\lambda}{4} \eta^{\mu\nu} \sigma^2(x) \ren{\phi^{2}}^{(1)}-{\frac{3\lambda^2}{2^8\pi^2} \eta^{\mu\nu}\sigma^4(x)},
 \end{split}
\end{align}
where we have defined 
\begin{align}
 M_E:=s_1(1-s_1)q_1^2+s_2(1-s_2)q_2^2-2s_1s_2(q_1\cdot q_2).
\end{align}

A direct computation is arduous and collinear divergences are always threatening; 
for a planar geometry, 
keeping the $\mathi \epsilon$ prescription one can introduce the following basic form factors
\begin{align}
 {\basicF}_0(q_1,q_2):&= q_1 \left(q_1 \log \left(\frac{q_1^2}{\left(q_1-q_2\right){}^2}\right)+q_2 \log \left(\frac{\left(q_1-q_2\right){}^2}{q_2^2}\right)\right),
 \\
 {\basicF}_1(q_1,q_2):&=q_1^4 \log \left(\frac{q_1^2}{\left(q_1-q_2\right){}^2}\right)+q_1q_2^3 \log \left(\frac{\left(q_1-q_2\right){}^2}{q_2^2}\right),
 \\
 {\basicF}_2(q_1,q_2):&=q_1^4 \log \left(\frac{q_1^2}{\left(q_1-q_2\right){}^2}\right)+q_1^2q_2^2  \log \left(\frac{\left(q_1-q_2\right){}^2}{q_2^2}\right),
 \\
 {\basicF}_3(q_1,q_2):&=\left(q_1^2-q_2 q_1-\mathi \epsilon_1\right) \left(q_1 q_2+\mathi \epsilon_2\right)\label{eq:basicF3},
\end{align}
in which $\epsilon_{1,2}$ are prescription parameters for the Feynman propagator.
Using them we may write a closed expression valid for a planar background field {that varies only in the $z$ direction}:
\begin{align}\label{eq:EM_order2_general}
\ren{\phi^2(z)}^{(2)} &= -\frac{\lambda^2}{2^4(2\pi)^{4}} \text{Re} \Bigg\lbrace \int \dxd[q_1][]\dxd[q_2][] \,e^{-\mathi q_1  z}  \, \tilde {\sigma}^2(q_1-q_2)\, \tilde {\sigma}^2(q_2) \,\frac{\basicF_0}{\basicF_3}\Bigg\rbrace,
\\
\begin{split}
\ren{T{}^{\mu\nu}(z)}^{(2)}
&=\frac{\lambda\,\eta^{\mu\nu}}{8}  \sigma^2(z)\ren{\phi^{2}}^{(1)}
  -\frac{ \lambda^2}{3\cdot 2^6(2\pi)^{2}}  \eta^{\mu\nu}   \sigma^4(z)
  -\frac{\lambda^2}{3\cdot 2^5 (2\pi)^2}  \Big(\eta^{\mu}{}_{3}\eta^{\nu}{}_{3}+\frac{1}{2} \eta^{\mu\nu}\Big)  \sigma^4(z)
  \\
  &\hspace{1.5cm}-\frac{\lambda^2}{16(2\pi)^{4}} \operatorname{Re}\Bigg\lbrace\int \dx[q_1]\dx[q_2]\,\frac{e^{-\mathi q_1z}}{\basicF_3}\, \tilde\sigma^2(q_1-q_2) \,\tilde \sigma^2(q_2) 
  \\
  &\hspace{4cm} \times\left[\Big(\eta^{\mu}{}_{3}\eta^{\nu}{}_{3}+\frac{1}{2} \eta^{\mu\nu}\Big) \left(-\frac{1}{3}{ \basicF_1} +\frac{1}{2} {\basicF_2}\right)
  +\frac{\eta^{\mu\nu}}{12}{\basicF_1}\right]\Bigg\rbrace,
\end{split}
 \end{align}
 in which we are ommiting the variables $(q_1,q_2)$  in the form factors $\basicF_i$.
{A direct computation shows the conservation law \eqref{eq:cons sem} is satisfied at second perturbative order.}

If one considers time-independent backgrounds the expressions become more tractable
than in the general case. 
In particular, the total energy is represented by the following simple formula:
\begin{align}
\begin{split}
\label{ec:E_orden_2}
     E_{\text{ren}}^{(2)}  &= \frac{\lambda^2}{2^6 (2 \pi)^{5}}   \operatorname{Re} \bigg[ \int \dxd[\alls[q_1]][3]\, \tilde\sigma^2(\alls[q]_1)\,\tilde \sigma^2(-\alls[q]_1) 
    \,\log\left(\frac{\alls[q]_1^2}{\mu^2}\right)\bigg],
\end{split}
\end{align}
where the Fourier transform evaluated at spatial coordinates implies omitting time variables, i.e.
\begin{align}
 \tilde \sigma^2(\alls[q]):=\int \dxd[\alls[x]_{1}][3]\, e^{\mathi\, \alls[q] \cdot \alls[x]_1} \,  \sigma^2(\alls[x]_1).
\end{align}

\subsection{Boundary divergences of $\ren{\phi^2}$ for a barrier}\label{sec:divergences_phi2}
As explained in Sec. \ref{sec:perturbative}, even after the appropiate renormalization
procedure has been carried out, both $\ren{\phi^2}$ and $\ren{T^{\mu\nu}}$
display divergences at the points where the background field $\sigma$ is discontinuous. 
Employing the perturbative formalism that was developed in the preceding sections we can 
unravel the precise structure of these divergences. We will call them ``boundary divergences'' or ``surface divergences'',
as a way to distinguish them from the divergences that require renormalization, which will be called bulk divergences.

First of all, we will consider a barrier of height $\Delta \sigma$ depending on only just one spatial coordinate
\begin{align}\label{eq:barrier}
 \sigma^2_b(z):= \Delta \sigma^2 \left(\Theta(z-a)-\Theta(z-b)\right),
\end{align}
where $\Theta(z)$ is the Heaviside function, and we will refer to $a$ and $b$ as boundaries. Its Fourier transform reads
\begin{align}\label{eq:transform_barrier}
\begin{split}
 \tilde\sigma^2_b(q)
&=  \Delta\sigma^2 \frac{\mathi\left( e^{\mathi a q}-e^{\mathi b q}\right)}{q} .
\end{split}
\end{align}
From this expression one can already appreciate why divergences will occur in $\ren{\phi^2}^{(1)}$
for such a background: 
the convergence for large momenta is only conditionally guaranteed by the oscillatory exponentials.
In other words, at those points where the exponents cancel, mild divergences should be present. 
Indeed, this can be confirmed replacing $\sigma_b$ in expression \eqref{eq:phi2_order1},
as done in \cite{Mazzitelli:2011st}:
\begin{align}\label{eq:surface_divergence_phi_order1}
\begin{split}
 \ren{\phi^2_{b}}^{(1)}
 &= \frac{\lambda \Delta\sigma^2 }{2^5 \pi^2} \Bigg\lbrace -\text{sign}(z-a) \Big[ \gamma+\log \big( \mu |z-a|\big)\Big]+\text{sign}(z-b) \Big[ \gamma+\log \big( \mu |z-b|\big)\Big]\Bigg\rbrace,
\end{split}
\end{align}
where $\gamma$ is the Euler-Mascheroni constant. Even if this expression is divergent at the boundaries, 
it is local, in the sense that it only depends on the information of the local jump, 
and integrable, so that one is able to define  its mean value over any desired region in space.

One important thing to notice is that, if $\sigma^2$ 
or its derivatives have a finite number of discontinuities\footnote{Additional divergences may occur in cases where the background field starts oscillating unconstrainedly.},
the only type of divergences present in $\ren{\phi^2}^{(1)}$ are those in Eq. \eqref{eq:surface_divergence_phi_order1}. 
Indeed, if the discontinuities appear only in the derivatives of $\sigma^2$, 
then the Fourier transform will contain additional powers of the momentum that will guarantee a non-conditional convergence.

Analogously, if one considers the second- or higher-perturbative orders of $\ren{\phi^2}$, 
a dimensional argument shows that for large momenta the integrand should behave as a power that provides convergence of the integral,
cf. Eq. \eqref{ec:phi_orden_n}.

At this point, the educated reader may be worried about the IR and collinear divergences
that we have mentioned in Sec. \ref{sec:perturbative}.  
They will appear in higher order computations since we are dealing with massless fields; 
an appropriate regulator should thus be used, or at least the $\mathi \epsilon$ prescription from the Wick rotation should be kept 
(see a related discussion for the {\se} tensor in App. \ref{app:EM_order1}). 
They will also appear in our first order contribution only if the $\sigma^2$ profile decays too slowly at infinity, 
as is the case of a step function.

\subsection{Divergences in the  stress-energy tensor for a barrier}\label{sec:divergences_EM}

\subsubsection{First-order computation for a barrier}\label{sec:divergences_EM_order1}

Consider now the first-order expression \eqref{eq:EM_order1} for $\ren{T^{\mu\nu}}$,
{focusing for the time being on the non-local contribution.}
If we naively replace the background field with $\sigma^2_b$, then we end up with a formally divergent expression,
to which a meaning should be ascribed:
\begin{align}
\begin{split}\label{eq:Tmunu_order1_step}
 \ren{T^{\mu\nu}_b}^{(1)}
 &= \frac{\lambda\,\Delta\sigma^2}{48(2\pi)^3} \Big( \eta^{\mu\nu} + \delta^{\mu}{}_{3}\delta^{\nu}{}_{3} \Big) \text{Im} \Bigg\lbrace\int \dx[q_1] e^{-\mathi q_1 z} \left( e^{\mathi a q_1}-e^{\mathi b q_1}\right) q_1 \log\left( \frac{q_1^2}{\mu^2}\right)\Bigg\rbrace.
 \end{split}
\end{align}
In App. \ref{app:EM_order1} we show that this expression is well-defined in the sense of distributions, 
which is the natural language of quantum field theory (see for example \cite{Estrada:2012yn} or \cite{Ashtekar:2021dab} for a recent discussion in astrophysics).
In this section we will follow a  physical approach, introducing an exponential cutoff $c>0$ in the Fourier transform,
\begin{align}
\begin{split}
 \ren{T^{\mu\nu}_b}^{(1)}
 &=  \frac{\lambda\,\Delta\sigma^2}{48(2\pi)^3} \Big( \eta^{\mu\nu} + \delta^{\mu}{}_{3}\delta^{\nu}{}_{3} \Big) \,  \text{Im} \Bigg\lbrace\int \dx[q_1] e^{-\mathi q_1 z-c|q_1|} \left( e^{\mathi a q_1}-e^{\mathi b q_1}\right) q_1 \log\left( \frac{q_1^2}{\mu^2}\right)\Bigg\rbrace,
 \end{split}
\end{align}
which is tantamount to saying that we have smoothed the discontinuity in the background field. 
A straightforward computation gives
\begin{align}
\begin{split}\label{eq:EM_order1_exponential}
 \ren{T^{\mu\nu}_b}^{(1)}
 &= - \frac{\lambda\,\Delta\sigma^2}{48(2\pi)^3} \Big( \eta^{\mu\nu} + \delta^{\mu}{}_{3}\delta^{\nu}{}_{3} \Big) \,  \text{Im} \Bigg\lbrace 
  \frac{\log \left({\mu ^2}\right)+2 \big[\log (\mathi a-\mathi z+c)+\gamma -1\big]}{(-a+\mathi c+z)^2}
  \\
  &\hu -\frac{\log \left({\mu ^2}\right)+2 \big[\log (-\mathi a+\mathi z+c)+\gamma -1\big]}{(a+\mathi c-z)^2}
  \\
  &\hu\hu+\frac{\log \left({\mu ^2}\right)+2 \big[\log (-\mathi b+\mathi z+c)+\gamma -1\big]}{(b+\mathi c-z)^2}
  \\
  &\hu\hu\hu-\frac{\log \left({\mu ^2}\right)+2 \big[\log (\mathi b-\mathi z+c)+\gamma -1\big]}{(-b+\mathi c+z)^2} 
 \Bigg\rbrace.
\end{split}
\end{align}
Notice first of all that due to the tensorial structure the $\ren{T^{33}_b}^{(1)}$ vanishes; 
the only components that survive are the diagonal terms in the other directions.  
Second, \eqref{eq:EM_order1_exponential} means that, 
as we approach the barrier profile by taking $c\to 0^+$, $\ren{T^{\mu\nu}_b}^{(1)}$ should display a bump that resembles a 
divergence at the boundaries ($z_0$ may be either $a$ or $b$ in the following formula):
\begin{align}\label{eq:EM_order1_div}
 \ren{T^{\mu\nu}_b}^{(1)}(z) \overset{z\to z_0}{\sim} \text{sign}(z-z_0)\,(z-z_0)^{-2} \Big( \eta^{\mu\nu} + \delta^{\mu}{}_{3}\delta^{\nu}{}_{3} \Big).
\end{align}

\subsubsection{Second-order computation for a barrier}\label{sec:divergences_EM_order2}
The second-order contribution to the {\se} tensor share some similarities with 
the first-order computation of $\ren{\phi^2}$. 
Indeed, a power counting argument in \eqref{eq:EM_order2} shows that the integrals involved  in the computation
are conditionally-convergent in the UV as long as we are not evaluating the expressions at the boundaries; 
at those points, the oscillatory behaviour may disappear and a mild divergence shoud then occur.

 As a particular example, we may analize the divergent terms for the barrier in Eq. \eqref{eq:barrier}.
 It should be expected that divergences will arise unless some fortuitous cancellations take place, since already the first term, 
 i.e. the one involving $\ren{\phi^2}^{(1)}$, is divergent at the boundary.
 We leave the lengthy computations to App. \ref{app:divergent_EM_order2}, simply stating the result:
 \begin{align}\label{eq:EM_order2_div}
 \begin{split}
\ren{T_b{}^{\mu\nu}}^{(2)}\Big\vert_{z\to a}
 &=
-\frac{\lambda^2}{3\cdot2^4(2\pi)^{2}} (\Delta\sigma^2)^2\log(|z-a|) \Big( \eta^{\mu\nu} + \delta^{\mu}{}_{3}\delta^{\nu}{}_{3} \Big) +\cdots.
 \end{split}
\end{align}
As was the case described in Sec. \ref{sec:divergences_EM_order1} for the first-order contributions, 
the tensorial structure implies that $\ren{T{}^{33}_b}^{(2)}$  is finite, 
while the remaining diagonal components of the {\se} tensor will display a divergence.
In this case, it is an integrable logarithmic one
and it is of local nature, depending only on the discontinuity of the background field at the corresponding boundary.


\section{The adiabatic approach and planar inhomogeneities}\label{sec:adiabatic}

Up to this point we have shown how to to compute physical quantities in a perturbative expansion in powers of $\sigma^2$.
It is instructive  to compare them with the results obtained in other approximations, performing thus a cross-check. 
In this section we will employ an adiabatic- or WKB-type approach, in which instead of expanding in powers of $\sigma^2$ 
one performs an expansion in the number of derivatives acting on the background field.
Our main goal is to confirm the results of the precedent section regarding the divergences for discontinuous backgrounds.

It will prove useful to introduce a special notation. We will focus on planar inhomogeneities which depend on only one spatial coordinate, 
which without loss of generality  we choose to be $\perpe[x]$ (or simply $\perpez[x]$ for formulae involving only one coordinate). 
The spacetime coordinates perpendicular to this preferred direction will be denoted as $\para[x]$, 
while its spatial subset will be written as $\paras[x]$.
As we will see, in order to be able to perform an adiabatic expansion we will need to work with an Euclidean signature; we will thus first show 
how the Euclideanization of our theory proceeds.

\subsection{The stress-energy tensor in terms of the Euclidean propagator}\label{sec:EM_euclidean}
Since we have shown that all the relevant quantities can be written in terms of Feynman's propagator \eqref{ec:prop_def},
we begin by studying its alternative Euclidean expression. 
As a first step, we can Fourier transform it in the directions perpendicular to $\perpe[x]$:
\begin{align}\label{eq:HF_fourier_transform}
 G_F(x,y) &=\int \dk[\omega] \dkd[\paras[k]][\dime-2] e^{-\mathi \omega (x^{0}-y^{0})+\mathi \paras[k]\cdot (\paras[x]-\paras[y]) } \mathcal{G}(\omega,\paras[k];\perpe[x],\perpe[y]).
\end{align}
Imposing the fact that the background field depends on just the coordinate $\perpe[x]$,
the partially Fourier transformed propagator $\mathcal{G}$ (usually called reduced Green function) should satisfy the equation\footnote{We are setting $\lambda=1$ with respect to the previous sections. In writing $\perpe[\partial]_x$, we mean the partial derivative in the third direction of the coordinate $x$.}
\begin{align}
 \left(-\omega^2+{\paras[k]}^2-\left(\perpe[\partial]_{x}\right)^2+\frac{\sigma^2(\perpe[x])}{2}\right)\mathcal{G}(\omega,\paras[k];\perpe[x],\perpe[y])&=-\delta(\perpe[x],\perpe[y]).
\end{align}
If we perform a rotation to Euclidean space, i.e. $\omega_M\to \mathi \omega_E$, we obtain that the Euclidean propagator is a solution of the following differential equation:
\begin{align}
  \left(\omega_E^2+{\paras[k]}^2-\left(\perpe[\partial]_{x}\right)^2+\frac{\sigma^2(\perpe[x])}{2}\right)\mathcal{G}_E(\omega_E,\paras[k];\perpe[x],\perpe[y])&=-\delta(\perpe[x],\perpe[y]). \label{eq:eq_euclidean_gf}
\end{align}

In order to compute the propagator, instead of departing from \eqref{eq: TuvG} 
we will use an equivalent expression where a point-splitting is kept until the end of the computation:
\begin{align}\label{eq:point_splitting_Tmunu}
 \begin{split}
\langle T^{\mu\nu}\rangle(x)
 &=\left(\frac{1}{2}g^{\mu\nu}\left[\partial_{x,\alpha}\partial_y^{\alpha}-\frac{\sigma^2(\perpe[x])}{2}\right]-\partial_x^\mu \partial_{y}^{\nu}\right) 
 \\
 &\hu\times\int \dk[\omega] \dkd[\paras[k]][\dime-2] e^{-\mathi \omega (x^{0}-y^{0})+\mathi \paras[k]\cdot (\paras[x]-\paras[y]) } \text{Im}\Bigg\lbrace \mathcal{G}(\omega,\paras[k];\perpe[x],\perpe[y]) \Bigg\rbrace\Bigg\vert_{x=y}.
\end{split}
\end{align}
Keeping track of the Euclideanization also in the coordinates, Eq. \eqref{eq:point_splitting_Tmunu} becomes
\begin{align}\label{eq:euclidean_Tmunu}
 \begin{split}
&\vev{T^{\mu\nu}}(\mathi x_E^0,\alls)
 =\int \dk[\omega_E] \dkd[\paras[k]][\dime-2] 
 \\
 &\times \left(\frac{1}{2}g^{\mu\nu}\left[\omega_E^2+\paras[k]{}^2+\perpe[\partial]_{x}\perpe[\partial]_{y}+\frac{\sigma^2(\perpe[x])}{2}\right]+\partial'{}_{x_E}^\mu \partial'{}_{y_E}^{\nu}\right) \mathcal{G}_E(\omega_E,\paras[k];\perpe[x],\perpe[y])\big\vert_{x=y}
\end{split}
\end{align}
in terms of the formal vectors
\begin{align}
\begin{split}
 \partial'{}_{x_E}^\mu :&= (\omega_E, \mathi \paras[k],\perpe[\partial]_{x}),
 \\
 \partial'{}_{y_E}^\mu :&= (-\omega_E, -\mathi \paras[k],\perpe[\partial]_{y}).
\end{split}
\end{align}

We can further simplify this expression, taking into account that $\mathcal{G}_E$ must be invariant in the 
$(\dime-1)$-dimensional space $(\omega_E,\paras[k])$; performing the corresponding angular integration we find 
the desired expression,
\begin{align}\label{eq:euclidean_Tmunu_tensorial}
 \begin{split}
\vev{T^{\mu\nu}}(\mathi x_E^0,\alls)
 &=\frac{S_{\dime-2}}{2\pi^2}\int \dx[{\para[k]}](\para[k])^{\dime-2} \Bigg\lbrace\frac{1}{2}g^{\mu\nu}\left[\Big(1-\frac{H_{\dime-2}}{S_{\dime-2}}\Big)\para[k]{}^2+\perpe[\partial]_{x}\perpe[\partial]_{y}+\frac{\sigma^2(\perpe[x])}{2}\right]
 \\
&\hu\hu\hu+ \delta^{\mu}{}_{3}\delta^{\nu}{}_{3} \Big(\perpe[\partial]_{x}\perpe[\partial]_{y} -\frac{H_{\dime-2}}{S_{\dime-2}} \para[k]^2\Big)\Bigg\rbrace \mathcal{G}_E(\omega_E,\paras[k];\perpe[x],\perpe[y])\big\vert_{x=y},
\end{split}
\end{align}
in terms of the $(n-1)$-sphere's hyper-area,
\begin{align}
 S_{n-1}:&=\frac{2\pi^{n/2}}{\Gamma\left(\frac{n}{2}\right)},
\end{align}
and the projection factor
\begin{align}
 \begin{split}
H_{n-1}:&=\int \dx[\Omega_{n-1}] \cos^2(\phi_1)
 =\frac{\pi^{n/2}}{ \Gamma \left(\frac{n}{2}+1\right)}.
 \end{split}
\end{align}

\subsection{The adiabatic technique and planar inhomogeneities}

Now that we have recast the relevant expressions in terms of the Euclidean Green function, we need to compute the latter.
In general, the homogeneous version of Eq. \eqref{eq:eq_euclidean_gf} will have two linearly independent solutions, which we call $f_{\pm}$:
\begin{align}\label{eq:adiabatic_eq}
 (-\partial^2+\omega^2(x)) f_{\pm}(x)=0,
\end{align}
with $\omega^2(x):=\omega_E^2+{\paras[k]}^2+\frac{\sigma^2(\perpe[x])}{2}$.
One can use them to construct the corresponding Green function
as dictated by the theory of Sturm--Liouville operators,
\begin{align}\label{eq:euclidean_GF_canonical}
 \mathcal{G}_E(\para[k];\perpe[x],\perpe[x'])&=
 \frac{1}{[f_+,f_-]} f_{+} (\perpe[x]_>)f_{-} (\perpe[x]_<),
\end{align}
where $[f,g]$ is the Wronskian\footnote{We are defining the Wronskian as usual, i.e. $[f,g](x):=f(x)g'(x)-f'(x)g(x)$. }  between $f$ and $g$; additionally, $\perpe[x]_>$ ($\perpe[x]_<$) is the greatest (smallest) of the two numbers $\perpe[x]$ and $\perpe[x']$. 

However, in practice it is not possible to obtain the functions $f_{\pm}$ explicitly. The adiabatic approach is a way to obtain their expansions in powers of the derivatives of $\sigma^2$.
In this framework, one begins by proposing the substitution
\begin{align}\label{eq:adiabatic}
 f_{\pm}(\perpe[x])\to\frac{e^{\mp \int \dx\, W(\perpe[x])}}{\sqrt{2 W(\perpe[x])}},
\end{align}
where $W(\perpe[x])$ is the new unknown function.
Then one can propose an expansion of $W(\perpe[x])$  in the number of derivatives and obtain its coefficients recursively.
In App. \ref{app:adiabatic_coefficients} we show the first coefficients of this expansion.

We will focus on the case of an arbitrary background field, apart from the fact that it is discontinuous only at two planes. 
These two planes will be defined by the equations\footnote{As said before, to simplify the notation we will write $\perpez[x]$ instead of $\perpe[x]$.} $\perpez[x]=a,b$.
A formalism has been developed in previous works
to deal with this problem \cite{Bao:2015, Li:2019ohr}.
In those articles it has been shown that $f_{\pm}$ can be obtained by appropriately gluing 
the solutions obtained in each single slab of space where the background $\sigma^2$ is continuous. 
In short, we call 
\begin{align}
 \sigma^2(\perpez[x]) =
 \begin{cases}
  \sigma^2_1(\perpez[x]) ,\quad \perpez[x]<a
  \\
  \sigma^2_2(\perpez[x]),\quad a<\perpez[x]<b
  \\
  \sigma^2_3(\perpez[x]),\quad \perpez[x]>b
 \end{cases},
\end{align}
so that the solutions to the homogeneous equation \eqref{eq:adiabatic_eq} with $\sigma_i$ as background are called $e_{i,\pm}$, $i=1,2,3$;
the global solutions (with $\sigma$ as background) are denoted as $e_{\pm}$.
We provide more details in App. \ref{app:planar_boundaries}.

\subsection{The divergences of the two-point function}\label{sec:phi2_adiabatic}

In order to simplify the discussion, we will choose the following convention 
to fix the constants in the indefinite integrals involved in the adiabatic expansion, cf. \eqref{eq:adiabatic}:
\begin{align}
\begin{split}\label{eq:e_convention_integral}
 e_{1,\pm}\equiv\frac{e^{\mp\int_a^{\perpez[x]} W_{\sigma_1}\dx[{\perpez[x]}] }}{\sqrt{2W_{\sigma_1}}}, 
 \,e_{2,\pm}\equiv\frac{e^{ \mp\int_a^{\perpez[x]} W_{\sigma_2} \dx[{\perpez[x]}]}}{\sqrt{2W_{\sigma_2}}},
 \,
 e_{3,\pm}\equiv\frac{e^{\mp\int_b^{\perpez[x]} W_{\sigma_3} \dx[{\perpez[x]}]}}{\sqrt{2W_{\sigma_3}}}.
\end{split}
\end{align}
Of course, these arbitrary constants involved in the WKB expansion will play no role in the Green function,
given that they will cancel out when dividing by the appropiate Wronskians.
However, if we consider the convention in \eqref{eq:e_convention_integral}, the coefficients $A_{\pm}$, $B_{\pm}$, $C_{\pm}$  and $D_{\pm}$ defined in App. \ref{app:planar_boundaries} simplify, 
since then the Wronskians $[e_{i,+}, e_{i,-} ]\equiv 1, \,i=1,2,3$. 
In particular, employing \eqref{eq:e_convention_integral} it is immediate to express the Wronskian $[e_{+},e_{-}]$
(which as in the Sturm--Liouville problems is constant) in terms of different coefficients:
\begin{align}\label{eq:wronskian}
 \begin{split}
[e_{+},e_{-}] &=  C_+ =(B_-A_+ -A_-B_+ )
 = D_-.
 \end{split}
\end{align}
Using this information we may write the Euclidean reduced Green function in the following form:
\begin{align}\label{eq:G_E_equal_points_stepwise}
 \mathcal{G}_E(\para[k];\perpez[x])
 &=
  \begin{cases}
  \frac{C_-}{C_+}e^2_{3,+}+ \frac{1}{2W_{\sigma_3}}, \quad \perpez[x]>b
  \\
  \frac{A_-A_+}{C_+}e^2_{2,+}+ \frac{B_-B_+}{C_+}  e^2_{2,-}+ \frac{(B_-A_++A_-B_+ )}{C_+} \frac{1}{2W_{\sigma_2}}, \quad a<\perpez[x]<b
  \\
  \frac{D_+}{C_+} e^2_{1,-}+\frac{1}{2W_{\sigma_1}},\quad \perpez[x]<a
\end{cases}.
\end{align}

At this point the intuition tells us which are the divergent terms that require
renormalization: they will come from the terms proportional to $(W_{\sigma_i})^{-1}$, 
because the remaining terms are exponentially damped for large parallel momenta
(see the first coefficients of the adiabatic expansion in App. \ref{app:adiabatic_coefficients}).
However, some fortuitous cancellations of the exponential factors may take place 
at the boundaries as we will see later.

Before analyzing the divergences, it is better to extract from the Wronskians the polinomial dependence in $W_{\sigma_i}$; 
operationally calling this action ``polynomial'', we introduce then the definition
\begin{align}
 \begin{split}
 g_{i,j}^{s_1s_2}:&=\text{polynomial}([ e_{i,s_1} , e_{j,s_2} ])
 \\
 &=\frac{2W_{\sigma_j}(\perpez[x])W_{\sigma_i}(\perpez[x]) \left(-s_2 W_{\sigma_j}(\perpez[x])+ s_1 W_{\sigma_i}(\perpez[x])\right)+[ W_{\sigma_j}(\perpez[x]),\, W_{\sigma_i}(\perpez[x])] }{4 W_{\sigma_i}(\perpez[x]){}^{3/2} W_{\sigma_j}(\perpez[x]){}^{3/2}}.
\end{split}
\end{align}
In this way the coefficients are simplified to
\begin{align}\label{eq:coefficients_g}
 \begin{split}
 A_+&\to g_{3,2}^{+ -}(b) e^{\int_a^b W_{\sigma_2}},
 \quad
 B_+\to -g_{3,2}^{+ +}(b)e^{-\int_a^b W_{\sigma_2}},
 \\
 A_-&\to -g_{2,1}^{- -}(a),
  \quad
  B_-\to g_{2,1}^{+ -}(a),
\\
 C_+&\to g_{2,1}^{+ -}(a) g_{3,2}^{+ -}(b)  e^{\int_a^b W_{\sigma_2}}+g_{1,2}^{- -}(a) g_{3,2}^{+ +}(b)  e^{-\int_a^b W_{\sigma_2}},
 \\
 C_-&\to -g_{2,1}^{- -}(a) g_{2,3}^{+ -}(b)e^{-\int_a^b W_{\sigma_2}}-g_{3,2}^{- -}(b) g_{2,1}^{+ -}(a)e^{\int_a^b W_{\sigma_2}},
 \\
 D_+&\to -g_{2,1}^{+ +}(a) g_{3,2}^{+ -}(b)  e^{\int_a^b W_{\sigma_2}}-g_{3,2}^{+ +}(b) g_{1,2}^{+ -}(a)  e^{-\int_a^b W_{\sigma_2}},
 \\
 D_-&\to g_{2,1}^{- -}(a) g_{2,3}^{+ +}(b)e^{-\int_a^b W_{\sigma_2}}+g_{2,1}^{+ -}(a) g_{3,2}^{+ -}(b)  e^{\int_a^b W_{\sigma_2}}.
 \end{split}
\end{align}

\subsubsection{The renormalization}\label{sec:phi2_adiabatic_renormalization}
Now let us study the divergences that must be renormalized; we will call them bulk divergences. 
Employing the coefficients written in Eq. \eqref{eq:coefficients_g}, in the region $a<\perpez[x]<b$  one notices that
\begin{align}
 \begin{split}
  \frac{(B_-A_++A_-B_+ )}{C_+} \frac{1}{2W_{\sigma_2}}
  &=\frac{g_{2,1}^{+ -}(a)g_ {3, 2}^{+-} (b) e^{\int_a^b W_ {\sigma_ 2}}+g_ {2, 1}^{--} (a) g_ {3, 2}^{++} (b) e^{-\int_a^b W_ {\sigma_ 2}}}{g_ {2, 1}^{+-} (a) g_ {3, 2}^{+-} (b) e^{\int_a^b W_ {\sigma_ 2}} + 
 g_ {1, 2}^{--} (a) g_ {3, 2}^{++} (b) e^{-\int_a^b W_ {\sigma_ 2}}} \frac{1}{2W_{\sigma_2}}
 \\
 &\sim \frac{1}{2W_{\sigma_2}}+\cdots,
 \end{split}
\end{align}
up to exponentially decreasing functions for large momenta.
This kind of contribution is already explicit in the regions where $\perpez[x]<a$ or $\perpez[x]>b$.
An explicit computation in terms of the coefficients given in App. \ref{app:adiabatic_coefficients} 
gives an expansion in inverse powers of the parallel momenta,
\begin{align}\label{eq:1/2W}
 \frac{1}{2W_{\sigma_i}}
 &=\frac{1}{2 (\para[k]) }-\frac{\sigma_i(\perpez[x]){}^2}{8 (\para[k]) ^3}+\frac{-2 \sigma_i'(\perpez[x]){}^2-2 \sigma_i(\perpez[x]) \sigma_i''(\perpez[x])+3 \sigma_i(\perpez[x]){}^4}{32 (\para[k]) ^5}+\cdots.
\end{align}
This is enough to compute the bulk divergent terms of the two-point function;
indeed,
upon integration over the $\para[k]$-momentum variables we obtain
\begin{align}
 \begin{split}
  \vev{\phi^2(\perpez[x])}^{\rm WKB}&=\int \dkd[{\para[k]}][\dime-1] \frac{1}{2W_{\sigma_i}}
 \\
&=\frac{1}{\dime-4}\frac{\sigma^2(\perpez[x])}{16\pi^2} +\text{ finite terms}.
\end{split}
\end{align}
A direct computation shows that this coincides with both the SDWE adiabatic result in Eq. \eqref{eq: phi2 ad2}
and the perturbative one.
\subsubsection{The divergences at the boundaries}\label{sec:phi2_adiabatic_boundary}

For simplicity we will consider just the region where $\perpez[x]>b$; the remaining ones can be worked out in an analogous way. 
The contribution for large parallel momentum reads
\begin{align}\label{eq:phi2_boundary_div}
 \begin{split}
\frac{C_-}{C_+} e_{3,+´}^2
 &=-\frac{g_{2,1}^{- -}(a) g_{2,3}^{+ -}(b)e^{-2\int_a^b W_{\sigma_2}}+g_{3,2}^{- -}(b) g_{2,1}^{+ -}(a) }{ g_{2,1}^{+-}(a) g_{3,2}^{+-}(b) + g_{1,2}^{--}(a)g_{3,2}^{++}(b) e^{-2\int_a^b W_{\sigma_2}}}  \frac{e^{-2\int_b^{\perpez[x]} W_{\sigma_3}}}{{2 W_{\sigma_3}}}
 \\
 &=\Bigg\lbrace
 \frac{\sigma_3(b){}^2-\sigma_2(b){}^2}{16 (\para[k]) ^3}+\frac{\sigma_2(b) \sigma_2'(b)-\sigma_3(b) \sigma_3'(b)}{16 (\para[k]) ^4}
 \\
 &\hu+\frac{1}{32 (\para[k]) ^5} \Bigg[\sigma_2(b){}^2 \sigma_3(\perpez[x]){}^2-\sigma_3(b){}^2 \sigma_3(\perpez[x]){}^2-\sigma_2'(b){}^2+\sigma_3'(b){}^2
 \\
 &\hu-\sigma_2(b) \sigma_2''(b)+\sigma_3(b) \sigma_3''(b)+\sigma_2(b){}^4-\sigma_3(b){}^4
 \Bigg]+\cdots 
 \Bigg\rbrace e^{-2\int_b^{\perpez[x]} W_{\sigma_3}} +\cdots .
 \end{split}
\end{align}
The situation is now patent: 
the exponential decay is guaranteed for any $\perpez[x]\neq b$; however, when $\perpez[x] \to b^+$ 
the exponent vanishes and gives rise to divergences if the inverse powers of $\para[k]$ of the expression in \eqref{eq:phi2_boundary_div} are not large enough. 
Of course, in $D=4$ the boundary divergences of $\ren{\phi^2}$ involve only the $(\para[k])^{-3}$ contribution; 
we have written also the higher order contributions that will be relevant in the analysis of the {\se} tensor. 

At this point a direct computation shows the exact form of the boundary divergence:
\begin{align}
 \begin{split}
  \int_{\para[k]\geq1} \dkd[\para[k]][\dime-1] & \left[\frac{\sigma_3(b){}^2-\sigma_2(b){}^2}{16 (\para[k]) ^3}\right] e^{-2 (\perpez[x]-b) \para[k]{}}
  =- \frac{\Delta \sigma^2(b)}{32 \pi^2} ( \log (\perpez[x]-b)+\gamma )+ \mathcal{O}({\perpez[x]-b}),
 \end{split}
\end{align}
where $\Delta \sigma^2 (y)$ denotes the jump of the background field at y, i.e.
\begin{align}
 \Delta \sigma^2 (y)= \sigma^2 (y^+)-\sigma^2 (y^-).
\end{align}

Computing the remaining contributions, one obtains a result that coincides with the one obtained in  Eq. \eqref{eq:surface_divergence_phi_order1}.
Notice however that in this section our conclusion is not restricted to a given power in $\sigma^2$. 
Then one can conclude that the only boundary divergences present in $\ren{\phi^2}$ are all linear in $\sigma^2$.

\subsection{The divergences of the stress-energy tensor}\label{sec:EM_adiabatic}

Taking into account Eq. \eqref{eq:euclidean_Tmunu_tensorial}, 
the divergences' structure of the {\se} tensor can be analyzed in a manner analogous to that for $\vev{\phi^2}$.
The only difference is that we additionally need an expansion for the product of derivatives\footnote{There are additional terms involving ultralocal factors that vanish in dimensional regularization.} $e'_{+}(\perpez[x]) e'_{-} (\perpez[x])$.
The computation is straightforward, albeit lengthy; this can be appreciated already from its structure:
\begin{align}
 \begin{split}\label{eq:e'e'}
  \frac{1}{[e_+,e_-]} e'_+ e'_-
&= 
  \begin{cases}
  \frac{C_-}{C_+} \frac{\left(W_{\sigma_3}'(\perpez[x])+2 W_{\sigma_3}(\perpez[x]){}^2\right){}^2 }{8 W_{\sigma_3}(\perpez[x]){}^3} e^{-2 \int_b^{\perpez[x]} W_{\sigma_3} \, \dx[{\perpez[x]}]}
  \\
  \hu+ \frac{W_{\sigma_3}'(\perpez[x]){}^2-4 W_{\sigma_3}(\perpez[x]){}^4}{8 W_{\sigma_3}(\perpez[x]){}^3}, \quad \hu\hu\hu\perpez[x]>b
  \\[0.5cm]
  \frac{A_-A_+}{C_+} \frac{\left(W_{\sigma_2}'(\perpez[x])+2 W_{\sigma_2}(\perpez[x]){}^2\right){}^2 }{8 W_{\sigma_2}(\perpez[x]){}^3} e^{-2 \int_a^{\perpez[x]} W_{\sigma_2}(\perpez[x]) \, \dx[{\perpez[x]}]}
  \\
  \hu-\frac{B_-B_+}{C_+} \frac{\left(W_{\sigma_2}'(\perpez[x])-2 W_{\sigma_2}(\perpez[x]){}^2\right){}^2 }{8 W_{\sigma_2}(\perpez[x]){}^3} e^{2 \int_a^{\perpez[x]} W_{\sigma_2}(\perpez[x]) \, \dx[{\perpez[x]}]}
  \\
  \hu\hu+ \frac{(A_+B_- +A_-B_+)}{C_+}\frac{W_{\sigma_2}'(\perpez[x]){}^2-4 W_{\sigma_2}(\perpez[x]){}^4}{8 W_{\sigma_2}(\perpez[x]){}^3}, \quad a<\perpez[x]<b
  \\[0.5cm]
  \frac{W_{\sigma_1}'(\perpez[x]){}^2-4 W_{\sigma_1}(\perpez[x]´){}^4}{8 W_{\sigma_1}(\perpez[x]){}^3}
  \\
  \hu+\frac{D_+}{C_+} \frac{\left(W_{\sigma_1}'(\perpez[x])-2 W_{\sigma_1}(\perpez[x]){}^2\right){}^2 }{8 W_{\sigma_1}(\perpez[x]){}^3} e^{2 \int_a^{\perpez[x]} W_{\sigma_1}(\perpez[x]) \, \dx[{\perpez[x]}]},\quad \perpez[x]<a
\end{cases}.
 \end{split}
\end{align}

\subsubsection{The renormalization}\label{sec:EM_adiabatic_renormalization}
Eq. \eqref{eq:euclidean_Tmunu_tensorial} contains several factors that are not exponentially suppressed for large parallel momentum. 
The large-$\para[k]$ expansion of many of them have already been derived in Sec. \ref{sec:phi2_adiabatic_renormalization}.
The only new contribution of this type, inherited from expression \eqref{eq:e'e'}, can be expanded as 
\begin{align}
 \begin{split}
  \frac{W_{\sigma_i}'(\perpez[x]){}^2-4 W_{\sigma_i}(\perpez[x]){}^4}{8 W_{\sigma_i}(\perpez[x]){}^3}
  &= -\frac{(\para[k]) }{2}-\frac{\sigma_i(\perpez[x]){}^2}{8 (\para[k]) }+\frac{-4  \sigma_i'(\perpez[x]){}^2-4  \sigma_i(\perpez[x]) \sigma_i''(\perpez[x])+\sigma_i(\perpez[x]){}^4}{64 (\para[k]) ^3}+\cdots .
 \end{split}
\end{align}
Summing all the contributions, in dimensional regularization we obtain
\begin{align}\label{eq:bulk_div_EM}
 \begin{split}
 \vev{T^{\mu\nu}}^{\rm WKB}
 &=\frac{\left(\eta^{\mu\nu}+ \delta^{\mu}{}_3\delta^{\nu}{}_3\right)}{\dime-4}\frac{1}{3\cdot2^5(2\pi)^2} 
 \Big(4 (\sigma^2)''+3 \sigma^4\Big)+ \text{ finite terms},
 \end{split}
\end{align}
which agrees with our perturbative computation, as well as with the adiabatic approach in the SDWE framework, cf. \eqref{eq: Tuv div}.

\subsubsection{The boundary divergences}\label{sec:EM_adiabatic_boundary}
Boundary divergences arise as in the case of $\vev{\phi^2}$,
i.e. some exponentially decreasing factors that decrete the convergence of the integrals for large momenta may disappear at the boundaries.
As an example, consider the following term from expression \eqref{eq:e'e'}, for $\perpez[x]>b$:
\begin{align}\label{eq:e'e'_boundary}
 \begin{split}
   \frac{C_-}{C_+} &\frac{\left(W_{\sigma_3}'(\perpez[x])+2 W_{\sigma_3}(\perpez[x]){}^2\right){}^2 }{8 W_{\sigma_3}(\perpez[x]){}^3} e^{-2 \int_b^{\perpez[x]} W_{\sigma_3} \, \dx[{\perpez[x]}]}
   \\
   &=\Bigg\lbrace
   \frac{\sigma_3(b){}^2-\sigma_2(b){}^2}{16 (\para[k]) }+\frac{\sigma_2(b) \sigma_2'(b)-\sigma_3(b) \sigma_3'(b)}{16 (\para[k]) ^2}
   \\
   &\hu+\frac{1}{32 (\para[k]) ^3}
   \Big[-\sigma_2(b){}^2 \sigma_3(\perpez[x]){}^2+\sigma_3(b){}^2 \sigma_3(\perpez[x]){}^2-\sigma_2'(b){}^2
   \\
   &\hu+\sigma_3'(b){}^2-\sigma_2(b) \sigma_2''(b)+\sigma_3(b) \sigma_3''(b)+\sigma_2(b){}^4-\sigma_3(b){}^4+\cdots\Big]
   \Bigg\rbrace e^{-2 \int_b^{\perpez[x]} W_{\sigma_3} \, \dx[{\perpez[x]´}]} +\cdots.
 \end{split}
\end{align}
Although the exponent provides the necessary fast decay for large $\para[k]$, it happens only if $\perpez[x] \neq b$. 
Replacing in Eq. \eqref{eq:euclidean_Tmunu_tensorial} both the contributions analogue to \eqref{eq:e'e'_boundary} and the results of Sec. \ref{sec:phi2_adiabatic_boundary},
we finally obtain
\begin{align}\label{eq:EM_adiabatic_div}
\begin{split}
&\vev{ T^{\mu\nu}}_{\perpez[x] \to \perpez[x]_0}^{\rm WKB}= \frac{\Big( \eta^{\mu\nu} + \delta^{\mu}{}_{3}\delta^{\nu}{}_{3} \Big)}{3\cdot 2^4 (2 \pi) ^2 }   H(\perpez[x],\perpez[x]_0),
\end{split}
\end{align}
where $\perpez[x]_0=a,b$ and we have defined the scalar function
\begin{align}
\begin{split}
H(x,y)
:&=\text{sign}(x-y) \left[\frac{  \Delta \sigma^2(y) }{(x-y)^2}-\frac{\Delta \Big((\sigma^2)' \Big)(y)}{(x-y)}\right]
\\
&\hu+ \log (|y-x|) \left[- \Big[\Delta \left(\sigma^2\right)(y)\Big]^2-\text{sign}(x-y)\Delta\left( (\sigma^2)''\right)(y)\right].
\end{split}
\end{align}
In particular, if we restrict ourselves to the case of a barrier, then we reobtain the results \eqref{eq:EM_order1_div} and \eqref{eq:EM_order2_div}.
The importance of the expansion \eqref{eq:EM_adiabatic_div} resides in the following two facts: 
in the worst case divergences are of second order in powers of $\sigma^2$, 
so all of them can be studied by our perturbative expressions in Sec. \ref{sec:perturbative_examples},  
and they are of local nature, 
confirming that they will play no role in the computation of Casimir forces.


\section{Conclusions}\label{sec:conc}

We have employed a perturbative method, toghether with dimensional regularization and adiabatic renormalization,
to prove master formulas for a scalar model in the realm of generalized Lifshitz configurations.

First of all, we have provided a general (perturbative) proof of the validity of the principle of virtual work,
showing that in the time-independent situation one can indeed compute the Casimir force exerted on one body in two different ways: 
either by considering the change in the energy of the system after a virtual displacement of the body, or by computing the stresses acting on the latter, cf. Eq. \eqref{eq:pvw}.
The derivation is valid for arbitrary geometries and to all order in the perturbation.

A fundamental pillar that allowed the proof was the conservation law that the energy-stress tensor satisfies not only at the classical level, 
but also at the level of renormalized VEVs in the semiclassical theory (quantum for the $\phi$ field and classical for the background one), as is guaranteed by Eq.~\eqref{eq:cons sem}.
This is a highly nontrivial point, since in general the regularization and the renormalization process may break classical laws at any point, introducing the so-called quantum anomalies. 

We have also provided master expressions for the $n$-th perturbative order VEVs of the two-point function and the energy-stress tensor. 
In particular, we have shown that in the static case only $\vev{\phi}^{(n-1)}$ is required in order to compute the total energy of the system at order $n$.
Given that the complexity of the calculations increases with the order of the perturbation and is greater for $\vev{T^{\mu\nu}}$ than for $\vev{\phi^2}$, 
we believe that such a formula will be extremely useful
for evaluating the vacuum energy in concrete examples. 
Additionally, we have written explicit formulas for all the relevant VEVs at first and second perturbative order, 
having computed the relevant form factors for planar configurations.

With the help of those master formulas, 
we have analyzed in detail the divergences that appear both in $\vev{\phi^2}$ and $\vev{T^{\mu\nu}}$ 
as a consequence of discontinuities in a planar background field,
extending the results in Refs. \cite{Bao:2015, Parashar:2018pds}. 
Our computations show that their functional dependence on the background field is at most quadratic, while they are local.
These considerations have been confirmed by an alternative WKB-type approach, 
proving that they are not relevant in the computation of Casimir forces.
For the mathematically-oriented reader, we have also dedicated a section regarding their formal interpretation in terms of distribution.

It is important to mention that, contrary to the situation when other renormalization prescriptions are employed as in Ref. \cite{Estrada:2012yn}, we do not obtain a so-called pressure anomaly. 
Moreover, we do not find the analog of the van der Waals anomaly discussed in Ref. \cite{Efrat:2021avl}, which in our scalar model would consist in a violation of the semiclassical conservation equation for the energy-stress tensor.

In spite of the obtained results, there are still many open questions. 
The first one is related to the intrinsic character of the background field
in a given body and its sorroundings, and how they are
 affected by a displacement of the body.
Another interesting issue is whether our results regarding the surface divergencies can be extended to non-planar geometries,
either by considering the perturbative or the WKB-type approach.
These lines are currently being studied.

\section*{Acknowledgments}
The authors thank F. Schaposnik and H. Falomir for valuable discussions.
This research was supported by ANPCyT, CONICET, and UNCuyo. 
SAF is grateful to G. Gori and the Institut f\"ur Theoretische Physik, Heidelberg, for their kind hospitality.
SAF acknowledges support by UNLP, under project grant X909 and ``Subsidio a Jóvenes Investigadores 2019''.


\appendix
\section{First-order stress-energy tensor as a distribution}\label{app:EM_order1}
We have seen in Sec. \ref{sec:divergences_EM}
that, when we consider the first-order {\se} tensor for a barrier, 
 we obtain a formally divergent expression.
However, that expression is well-defined as a distribution.
As explained in \cite{Gelfand:1964_tomo1} (see also \cite{Falomir:metodos} for an introductory course),
we can interpret Eq. \eqref{eq:Tmunu_order1_step} as the Fourier transform of a distribution,
\begin{align}
 \begin{split}
  \int \dx[q_1] e^{-\mathi q_1 x} \left( e^{\mathi a q_1}-e^{\mathi b q_1}\right) q_1 \log\left( \frac{q_1^2+\mathi \epsilon}{\mu^2}\right)
  &=2 \mathcal{F}\Big[\alpha(q_1)\Big](a-x)-2 \mathcal{F}\Big[\alpha(q_1)\Big](b-x) ,
 \end{split}
\end{align}
where 
\begin{align}
 \alpha(q):&= q  \log\left( \frac{|q|}{\mu}\right).
\end{align}
To be more explicit, we may recast this expression in the notation of \cite{Gelfand:1964_tomo1} as
\begin{align}
 \begin{split}
\alpha(q)= q_+ \log(q_+)-q_- \log(q_-)-q\log(\mu),
\end{split}
\end{align}
defining the distributions 
\begin{align}
 x_{\pm}^{\lambda}&=\Theta(\pm x) |x|^{\lambda}.
\end{align}
Employing several identities that can be found in \cite{Gelfand:1964_tomo1} we end up with
\begin{align}
 \begin{split}\label{eq:transform_alpha}
  \mathcal{F}_{\rm Shilov}\Big[\alpha(x)\Big]
  &=-i \pi  \left(-2 (\log(\mu)+\gamma -1) \delta'(x)+\hat x^{-2}-2 x_-^{-2}\right) 
 \end{split}
 \end{align}
where both $x^{-2}$ and $ x_-^{-2}$ are themselves distributions that are defined by their actions on test functions $\phi$:
\begin{align}
 \label{eq:x-2} ( x^{-2},\phi)&=\int_0^{\infty} \dx\, x^{-2} \Bigg[\phi(x)+\phi(-x)-2\phi(0) \Bigg],
 \\
 (x_{-}^{-2},\phi) &= \int_0^{\infty} \dx \,x^{-2}\Big[\phi(- x)-\phi(0)+ x \phi'(0)\Theta(1-x)\Big].
\end{align}
The {\se} tensor can be readily obtained combining the previous equations:
\begin{align}
\begin{split}
 \ren{T_{\mu\nu}}^{(1)}
 &= \frac{\lambda\,(\delta^{3}{}_{\mu}\delta^{3}{}_{\nu}+ \eta_{\mu\nu})}{48(2\pi)^2} \Bigg\lbrace
 2  (-1+ \gamma+\log(\mu)) (\delta'(a-z)-\delta'(b-z))
 \\
 &\hu\hu-{(z-a)}^{-2}+ 2 {(z-a)_-}^{-2}
 +{(z-b)}^{-2}-2 (z-b)_-^{-2}
 \Bigg\rbrace.
 \label{eq:Tmunu_pert_1_final}
\end{split}
\end{align}
If one is interested just in mean values of $\ren{T_{\mu\nu}}^{(1)}$ over a finite region, 
then using \eqref{eq:Tmunu_pert_1_final} one obtains a finite number. 

\section{Second-order contributions in $\ren{T^{\mu\nu}}$ for a barrier and distributions}\label{app:divergent_EM_order2}
In general, Eq. \eqref{eq:EM_order2_general} involves regularized quadratic forms 
as given by the definition of $\basicF_3$ in \eqref{eq:basicF3}.
The mathematical theory has been extensively studied in \cite{Gelfand:1964_tomo1}; 
in this appendix we will follow a physicist approach, 
performing a change of variables that converts the quadratic forms into linear ones.
We can define the transformation 
\begin{align}\label{eq:p}
 (p_1,p_2):&=(q_1q_2,q_1(q_1-q_2)),
\end{align}
as well as its inverse, defining the functions $h^{\pm}(\cdot,\cdot)$:
\begin{align}
\begin{split}
\left(q_1,q_2\right)&=\pm \left( \sqrt{p_1+p_2}, \frac{p_1}{\sqrt{p_1+p_2}}\right)
\\
&=:h^{\pm}(p_1,p_2).
\end{split}
\end{align}
Taking into account the map of the domains we get
\begin{align}
 \begin{split}
  H:&=\int_{-\infty}^{\infty} \dx[q_1]\dx[q_2]\,  \frac{f(q_1,q_2)}{ \left[q_1 \left(q_1-q_2\right)-\mathi \epsilon _1\right] \left(q_1 q_2+\mathi \epsilon _2\right)}  
  \\
  &=\int_{-\infty}^{\infty}\int_{-p_1}^{\infty} \dx[p_2]  \dx[p_1]   \frac{\bar f\big(p_1,p_2\big)}{ \left[p_2-\mathi \epsilon _1\right] \left(p_1+\mathi \epsilon _2\right)}
\end{split}
\end{align}
for an arbitrary well-behaved function $f(\cdot,\cdot)$, if we use the additional definition
\begin{align}
 \bar{f}(p_1,p_2):=\frac{1}{2(p_1+p_2)} \big[f\big(h^+(p_1,p_2)\big)+f\big(h^-(p_1,p_2)\big)\big].
\end{align}
Using the Sokhotski–-Plemelj theorem to rewrite the $(p_i\pm\mathi \epsilon)^{(-1)}$ factors, we may further simplify this expression.
In particular, if the contributions corresponding to integrals of Dirac $\delta$ vanish (as is the case for the barrier), then we get
\begin{align}
 \begin{split}
  H\to H_{\rm PV,PV}
  &=\int_{0}^{\infty}\frac{\dx[p_1]}{p_1}\int_{0}^{p_1} \frac{\dx[p_2] }{p_2}  \Bigg\lbrace  \bar{f}(p_1,p_2)- \bar{f}(p_1,-p_2)+ \bar{f}(p_2,p_1) -  \bar{f}(-p_2,p_1) 
  \Bigg\rbrace.
  \end{split}
\end{align}
For a barrier it proves convenient to perform the rescalings $p_2\to p_1p_2$ and afterwards $p_1\to (1\pm p_2)p_1$, depending on whether the argument in $\bar f$ is $\pm p_2$.
The integral in $p_1$ can then be performed and the remaining integral in $p_2$ is convergent for $x\neq a,b$. The contributions that are divergent at the boundaries can be isolated.
To sketch the kind of computations involved for a barrier, consider the following example:
\begin{align}
\begin{split}
 &\left.\int \dx[q_1]\dx[q_2]\, \tilde\sigma_b^2(q_1-q_2)\tilde \sigma_b^2(q_2) e^{-\mathi q_1 x} \frac{{\basicF_1}}{\basicF_3} \right\vert_{\rm div}
 =8\int_0^1 \frac{\dx[p_2]}{ p_2^2 \left(p_2^2-1\right)} 
 \\
 &\hu\times
 \Big[p_2^4 \left(\log \left(1-p_2\right)-2 \log \left(p_2\right)+\log \left(p_2+1\right)\right)-2 p_2^3 \left(\log \left(1-p_2\right)-\log \left(p_2+1\right)\right)
 \\
 &\hu\hu+2 p_2 \left(\log \left(1-p_2\right)-\log \left(p_2+1\right)\right)-\log \left(1-p_2\right)-\log \left(p_2+1\right)\Big]
 \\
 &\hspace{6cm}\times  \log \left(\frac{\left| a-x\right|  \left| b-x\right| }{\left| b-x+(x-a) p_2\right|  \left| a-x+(x-b) p_2\right| }\right).
\end{split}
\end{align}
The divergence thefore arises either directly from a $\log(|x-y|)$ term, or from a term that renders the integrand singular when $x=y$.
Both of them can be easily handled and the sum over all the form factors results in Eq. \eqref{eq:EM_order2_div}.

\section{Adiabatic coefficients}\label{app:adiabatic_coefficients}
If we try to solve Eq. \eqref{eq:adiabatic_eq} by using Eq. \eqref{eq:adiabatic} toghether with the \emph{ansatz} 
\begin{align}
 W(x)=\sum_{j=0}^{\infty} W_{j}(x),
\end{align}
where $W_j$ is a term of adiabatic order $j$,
then we find that the coefficients with odd $j$ vanish.
The first ones for even $j$ are given by
\begin{align}
\begin{split}
 W_0(x)&=\omega (x),
  \\
 W_2(x)&= -\frac{3 \omega '(x)^2}{8 \omega (x)^3}+\frac{\omega ''(x)}{4 \omega (x)^2},
 \\
 W_4(x)&= \frac{\omega ^{(4)}(x)}{16 \omega (x)^4}-\frac{13 \omega ''(x)^2}{32 \omega (x)^5}-\frac{297 \omega '(x)^4}{128 \omega (x)^7}-\frac{5 \omega ^{(3)}(x) \omega '(x)}{8 \omega (x)^5}+\frac{99 \omega '(x)^2 \omega ''(x)}{32 \omega (x)^6}.
 \end{split}
 \end{align}

\section{Green function for planar boundaries}\label{app:planar_boundaries}

In this appendix we review the results of \cite{ Li:2019ohr}.
If we have a discontinuous background field, 
we can obtain the solutions to the inhomogeneous eq. \eqref{eq:eq_euclidean_gf} 
by gluing toghether the solutions to the inhomogeneous problem in each slab, which we call $e_{\pm,i}$.
As customary, there will exist two solutions; we will call them $e_{\pm}$, according to whether they decay fast enough at $\pm\infty$. 
If we ask $e_{\pm}$ and their first derivatives  to be continuous at $\perpez[x]=a,b$, the expansion read as follows:
\begin{align}
 e_+(\perpez[x]):=
 \begin{cases}
  e_{3,+},\quad \perpez[x]>b
  \\
  A_+e_{2,+}+B_+ e_{2,-}, \quad a<\perpez[x]<b
  \\
  C_+e_{1,+}+D_+ e_{1,-}, \quad \perpez[x]<a
 \end{cases},
\end{align}
\begin{align}
 e_-(\perpez[x]):=
 \begin{cases}
  C_-e_{3,+}+D_- e_{3,-}, \quad \perpez[x]>b
  \\
  A_-e_{2,+}+B_- e_{2,-}, \quad a<\perpez[x]<b
  \\
  e_{1,-},\quad \perpez[x]<a
\end{cases},
\end{align}
in terms of the coefficients
\begin{align}
 \begin{split}
A_+&= \frac{[ e_{3,+} , e_{2,-} ] ( b )}{[ e_{2,+} , e_{2,-} ] ( b )},\; A_-= -\frac{[ e_{2,-} , e_{1,-} ] ( a )}{[ e_{2,+} , e_{2,-} ] ( a )},\;
B_+= -\frac{[ e_{3,+} , e_{2,+} ] ( b )}{[ e_{2,+} , e_{2,-} ] ( b )},\;B_-= \frac{[ e_{2,+} , e_{1,-} ] ( a )}{[ e_{2,+} , e_{2,-} ] ( a )},
\\
C_+&= \frac{A_+ [ e_{2,+} , e_{1,-} ] ( a )-B_+ [ e_{1,-} , e_{2,-} ] ( a )}{[ e_{1,+} , e_{1,-} ] ( a )}, 
\;C_-= -\frac{A_- [ e_{2,+} , e_{3,-} ] ( b )+B_- [ e_{2,-} , e_{3,-} ] ( b )}{[ e_{3,+} , e_{3,-} ] ( b )},
\\
D_+&= \frac{B_+ [ e_{1,+} , e_{2,-} ] ( a )-A_+ [ e_{2,+} , e_{1,+} ] ( a )}{[ e_{1,+} , e_{1,-} ] ( a )},
\;D_-= -\frac{B_- [ e_{3,+} , e_{2,-} ] ( b )+A_- [ e_{3,+} , e_{2,+} ] ( b )}{[ e_{3,+} , e_{3,-} ] ( b )}.
\end{split}
\end{align}
and the Wronskians $[f,g](x):=f(x)g'(x)-f'(x)g(x)$. The main difference 
with the results in \cite{ Li:2019ohr} resides in the fact that our Wronskians are the usual ones.


\bibliography{bibliografia}

\end{document}